\documentclass[aps,nofootinbib,notitlepage,showpacs,longbibliography]{revtex4-1}

\usepackage[CJKbookmarks, pdftex, bookmarksnumbered, bookmarksopen, colorlinks, citecolor=magenta, linkcolor=blue]{hyperref}

\usepackage{amstext,amsmath,amssymb,amsfonts,bbm}
\usepackage[latin1]{inputenc}
\usepackage{graphicx}
\usepackage{epstopdf}
\usepackage{amsthm}
\usepackage{tocvsec2}
\usepackage{enumerate}
\usepackage{subfigure}
\usepackage{color}

\newcommand{\lyxdot}{.}
\usepackage{multirow} \usepackage{rotating}

\def\beq{\begin{equation}}
\def\be{\begin{equation}}
\def\ee{\end{equation}}
\def\bes{\begin{eqnarray}}
\def\ees{\end{eqnarray}}

\begin{document}

\title{Degrees of freedom in discrete geometry}

\author{Seramika Ariwahjoedi$^{1,3}$, Jusak Sali Kosasih$^{3}$, Carlo Rovelli$^{1,2}$, Freddy P. Zen$^{3}$\vspace{1mm}}

\affiliation{$^{1}$Aix Marseille Universit\'e, CNRS, CPT, UMR 7332, 13288 Marseille, France.\\
$^{2}$Universit\'e de Toulon, CNRS, CPT, UMR 7332, 83957 La Garde, France.\\$^{3}$Institut Teknologi Bandung, Bandung 40132, West Java, Indonesia.}

\begin{abstract} 

\noindent
Following recent developments in discrete gravity, we study geometrical
variables (angles and forms) of simplices in the discrete geometry point
of view. Some of our relatively new results include: new ways of writing
a set of simplices using vectorial (differential form) and coordinate-free
pictures, and a consistent procedure to couple particles of space,
together with a method to calculate the degrees of freedom of the
system of 'quanta' of space in the classical framework.
\end{abstract}

\maketitle

\section{Introduction}

Studies of discrete gravity arise in attempt to do numerical calculations
on general relativity, since the analytical solution to Einstein field
equation is usually hard to obtain, because in general, it requires
a solution to a coupled, second order, non-linear differential equation.
The first work in this field was started by Tullio Regge \cite{key-3.19},
as an attempt to rewrite the formulation of general relativity without
using coordinate systems. In this point of view, Regge calculus (or
discrete gravity) is a discrete approximation to general relativity.
Many developments and results on discrete gravity are obtained through
practical use of the theory, mainly through simulations on black holes
dynamics and gravitational waves \cite{key-3.9}.

In the other hand, loop quantum gravity (LQG) predicts the existence
of the 'atoms' of space \cite{key-3.1,key-3.2,key-3.3,key-3.4}, which
in the semi-classical limit, corresponds to quantum polyhedra \cite{key-3.5}.
In LQG point of view, discrete geometry is more fundamental than the
differential geometry picture, which means at the quantum scale, space
are predicted to be formed by discrete 'atoms' of space \cite{key-3.4}.
The continuous, smooth differential geometry is obtained only in an
asymptotical limit of the theory. Specifically, discrete geometry
is the \textit{mesoscopic}, or the \textit{semi-classical limit} of
LQG, obtained by taking the \textit{spin-number} $j$ (which is responsible
to the size of the quanta of space) to be large: $j\rightarrow\infty$
\cite{key-8.1,key-1.13,key-1.14}. Meanwhile, classical general relativity
is the \textit{classical 'continuum' limit} of the theory, obtained
by taking both the spin number and \textit{number of quanta} $n$
(which is responsible to the number of degrees of freedom) to be
large: $n,j\rightarrow\infty$. \cite{key-3.19,key-3.23,key-3.31,key-3.20,key-7.21}.
The latest result on the asymptotical limit of LQG can be found elsewhere
\cite{key-8.1,key-1.13,key-7.15,key-1.15,key-1.16}.

An important principle in general relativity is the\textit{ general
covariance} principle: every physical formulation must be invariant
under diffeomorphism/local coordinate transformation \cite{key-3.35}. This
principle is important because the formulation of classical general
relativity is written in a vectorial (tensorial) form. To get rid
of this, Regge reformulated general relativity without using \textit{any}
coordinate system, that is, by using \textit{scalars}, i.e., the area-angles
variables. The discrete structure of the theory allows him to write
GR free from coordinates \cite{key-3.19}. The discreteness of space
is also, naturally, compatible with \textit{background independence},
a fundamental principle adopted by many conservative theories of gravity
\cite{key-3.35,key-3.34}. 

Moreover, in LQG, it is important to be able to count the degrees
of freedom in a set of quanta of space. Specifically, there exist
a technical problem concerning the difference in the calculation of
the degrees of freedom from twisted geometry and Regge discrete geometry
\cite{key-5.2,key-5.3,key-5.4}. The exact number of degrees of freedom
of a set of simplices is crucial in proposing a classical coarse-graining
procedure, which is important to obtain the classical limit, in particular
\cite{key-1.13,key-1.16,key-4.3}. Consequently, to obtain the number
of degrees of freedom of a set of coupled simplices describing a chunk
of space, a consistent procedure of coupling simplices is needed.

This article is an attempt to solve these problems. In Section II,
we study the discrete geometry without refering to any continuous,
smooth, differentiable theory as its origin. Here, we will write the
simplices in a vectorial picture, using differential forms. Section III,
which will be the main result in this work, is about the procedure
to write geometrical variables in a coordinate-free picture. In this section,
we give a consistent procedure of coupling simplices, together with
a way to calculate the degrees of freedom of the system of 'quanta'
of space. In the last section, we conclude our work.

\section{Discrete geometry}

In this section, we will study discrete geometry as a set of simplices,
connected to each other. The study of simplices, or polytopes in general,
had been developed in \cite{key-3.16,key-3.17,key-3.18}. But the
first attempt to apply discrete geometry to gravity, kinematically,
was done by Regge in \cite{key-3.19}, and then the dynamics in Ponzano-Regge
model \cite{key-3.20}. These works are developed in the second order
formulation of gravity in 3-dimension. An attempt to write discrete
gravity in first order formulation had been done in \cite{key-3.21}.
Moreover, a 4-dimensional, Lorentzian signature discrete gravity is
already studied in \cite{key-3.22}, known as the Barret-Crane model.

\subsection{Simplices and forms}

Let us first review the definition of \textit{$p$-simplex}. A $p$-simplex
is the simplest, flat, $p$-dimensional polytope embedded in an $n$-dimensional
space $\mathbb{R}^{n}$, with $n\geq p$. The reason of using simplices
is due to the fact that they are completely determined by their edges
\cite{key-3.23}. A $0$-simplex is simply a point, $1$-simplex is
a segment, $2$-simplex is a triangle, $3$-simplex is a tetrahedron,
and so on. A $p$-simplex is constructed from $\left(p+1\right)$
numbers of $\left(p-1\right)$-simplices. See FIG. 1. 
\begin{figure}[h]
\centerline{\includegraphics[scale=0.85]{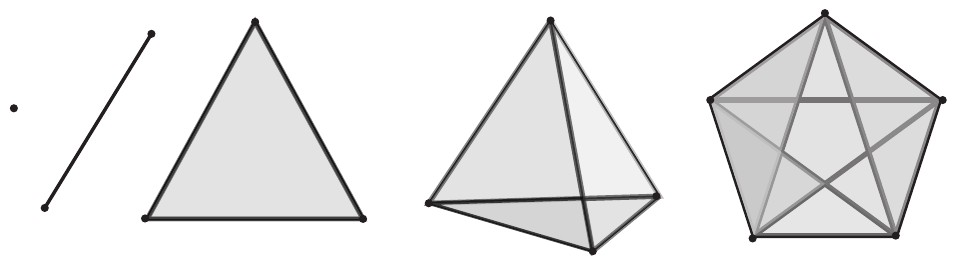}} \caption[Construction of a $p$-simplex.]{A $p$-simplex is constructed from $\left(p+1\right)$ numbers of
$\left(p-1\right)$-simplices: connecting two points construct a segment,
connecting three segments construct a triangle, connecting four triangles
construct a tetrahedron, connecting five tetrahedra construct a 4-simplex,
and so on. }
\end{figure}

We describe $p$-simplex using $p$-forms. A similar attempt had already
been studied in \cite{key-3.24}. Any \textit{$p$-form $\omega$}
can be written as:
\begin{equation}
\omega=\omega_{\left[a_{1}..a_{p}\right]}dx^{a_{1}}\wedge..\wedge dx^{a_{p}},\label{eq:3}
\end{equation}
with $\omega_{\left[a_{1}..a_{p}\right]}$ is an anti-symmetric tensor
of order $\tbinom{0}{p}$, given by:
\[
\omega_{\left[a_{1}..a_{p}\right]}=\frac{1}{p!}\delta_{a_{1}..a_{p}}^{b_{1}..b_{p}}\omega_{b_{1}..b_{p},}
\]
and $\delta_{a_{1}..a_{p}}^{b_{1}..b_{p}}$ is the generalized
Kronecker delta, given by:
\[
\delta_{a_{1}..a_{p}}^{b_{1}..b_{p}}=\det\left[\begin{array}{cccc}
\delta_{a_{1}}^{b_{1}} & \delta_{a_{2}}^{b_{1}} & \cdots & \delta_{a_{p}}^{b_{1}}\\
\delta_{a_{1}}^{b_{2}} & \ddots\\
\vdots\\
\delta_{a_{1}}^{b_{p}} & \delta_{a_{2}}^{b_{p}} & \cdots & \delta_{a_{p}}^{b_{p}}
\end{array}\right].
\]
A 1-form is simply a covariant vector:
\[
\mathbf{a}=a_{\mu}dx^{\mu},\qquad\mu=1,..,n.
\]
We can costruct $n$-form from several lower forms using the \textit{wedge
product} $\wedge$, for example, a 2-form from two 1-forms:
\begin{equation}
\mathbf{a}\wedge\mathbf{b}=\frac{1}{2}\left(a_{\mu}b_{\nu}-a_{\nu}b_{\mu}\right)dx^{\mu}\wedge dx^{\nu}.\label{eq:3.15}
\end{equation}
The space of $p$-forms over an $n$-dimensional space $\mathbb{R}^{n}$
is written as $\Omega^{p}\left(\mathbb{R}^{n}\right)$, together with
the wedge product $\wedge$, they form an \textit{exterior algebra}
over $\mathbb{R}^{n}$. See \cite{key-3.7,key-3.8,key-3.25} for details.
Since the space of $p$-forms $\Omega^{p}\left(\mathbb{R}^{n}\right)$
is a vector space satisfying vector space axioms, we can introduce
an inner product operation which give rise to a flat Euclidean metric
$g$ to the space $\Omega^{p}\left(\mathbb{R}^{n}\right)$. Using
this metric, we could obtain the contraction of two $p$-forms:
\begin{eqnarray}
g\left(\mathbf{a},\mathbf{b}\right) & = & g\left(a_{\left[\mu...\nu\right]}dx^{\mu}\wedge..\wedge dx^{\nu},b_{\left[\mu'...\nu'\right]}dx^{\mu'}\wedge..\wedge dx^{\nu'}.\right)\label{eq:3.16}\\
 & = & a_{\left[\mu...\nu\right]}b_{\left[\mu'...\nu'\right]}\varepsilon^{\mu..\nu}\varepsilon^{\mu'..\nu'},\nonumber 
\end{eqnarray}
with $\varepsilon^{\mu..\nu}\varepsilon^{\mu'..\nu'}$ comes from:
\begin{equation}
\varepsilon^{\mu..\nu}\varepsilon^{\mu'..\nu'}=g\left(dx^{\mu}\wedge..\wedge dx^{\nu},dx^{\mu'}\wedge..\wedge dx^{\nu'}\right),\label{eq:3aa}
\end{equation}
which is the tensor product of two Levi-Civita symbol in $n$-dimension.
The tensor product of two Levi-Civita symbol is a generalized Kronecker
delta, which can be obtained as:
\begin{equation}
\varepsilon^{\mu..\nu}\varepsilon^{\mu'..\nu'}=\det\left|\begin{array}{ccc}
\delta^{\mu\mu'} & \ldots & \delta^{\mu\nu'}\\
\vdots & \ddots & \vdots\\
\delta^{\nu\mu'} & \ldots & \delta^{\nu\nu'}
\end{array}\right|.\label{eq:3aaa}
\end{equation}

In the following subsection, we will give a sketch of the construction
of simplices using $p$-forms.

\subsubsection{2-simplex (triangle)}

A 1-form $\mathbf{l}$ can be interpreted geometrically as a segment
with length $\left|\mathbf{l}\right|$, from which we can construct
more complex geometries. A 2-simplex or a triangle can only be realized
in space with dimension $n\geq2.$ We can build a 2-simplex $\mathbf{a}$,
given two distinct 1-forms: $\left\{ \mathbf{l}_{1},\mathbf{l}_{2}\right\} \in T_{p}^{*}\mathcal{M}$
using the wedge product:
\[
\mathbf{a}=\frac{1}{2}\left(\mathbf{l}_{1}\wedge\mathbf{l}_{2}\right)\in T_{p}^{*}\mathcal{M}\otimes T_{p}^{*}\mathcal{M},
\]
where the components are given by (\ref{eq:3.15}). This can be interpreted
as the illustration shown in FIG. 2, with $\mathbf{a}$ is the triangle
constructed by segment $\mathbf{l}_{1}$ and $\mathbf{l}_{2}$. 
\begin{figure}[h]
\begin{centering}
\includegraphics[scale=0.85]{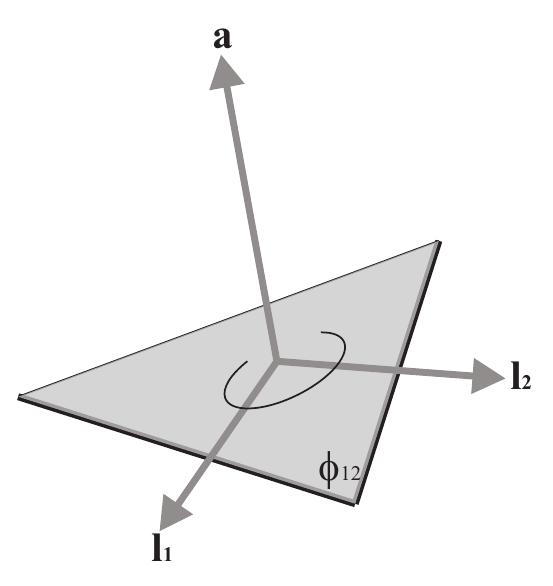}
\par\end{centering}

\caption{A triangle $\mathbf{a}$ can be obtained by wedge-producting two 1-forms:
$\mathbf{l}_{1}$ and $\mathbf{l}_{2}$. $\phi_{12}$ is the angle
between these 1-forms. }
\end{figure}
 Using the inner product on $\Omega^{2}\left(\mathbb{R}^{n}\right)$
defined in (\ref{eq:3.16}), we could obtain the norm of 2-form $\mathbf{a}$,
\[
\left|\mathbf{a}\right|=\sqrt{g\left(\mathbf{a},\mathbf{a}\right)},
\]
which is interpreted as the \textit{area of the triangle}.

We consider the \textit{boundary} of $\mathbf{a}$, constructed
from three 1-forms $\partial\mathbf{a}=\left\{ \mathbf{l}_{1},\mathbf{l}_{2},\mathbf{l}_{12}\right\} ,$
with $\mathbf{l}_{12}$ defined as:
\begin{equation}
\mathbf{l}_{12}\doteq\mathbf{l}_{1}+\mathbf{l}_{2},\label{eq:3.11}
\end{equation}
so that these set of 1-forms satisfy closure condition:
\begin{equation}
\mathbf{l}_{1}+\mathbf{l}_{2}-\mathbf{l}_{12}=0.\label{eq:3.12}
\end{equation}
Therefore, these set of forms satisfy the triangle inequality:
\begin{equation}
0\leq\left|\left(\mathbf{l}_{1}+\mathbf{l}_{2}\right)\right|\leq\left|\mathbf{l}_{1}\right|+\left|\mathbf{l}_{2}\right|,\label{eq:triequal}
\end{equation}
since metric $g$ on $\Omega^{2}\left(\mathbb{R}^{n}\right)$ is Riemannian.
Their norms can be interpreted as the \textit{length of segment} of
a closed triangle. The boundary $\partial\mathbf{a}$ is an example
of a \textit{subspace} of $\mathbf{a}.$

\subsubsection{3-simplex (tetrahedron)}

A 3-simplex or a tetrahedron can only be realized in space with dimension
$n\geq3.$ To build a 3-simplex $\mathbf{v}$, we need three distinct
1-forms: $\left\{ \mathbf{l}_{1},\mathbf{l}_{2},\mathbf{l}_{3}\right\} \in T_{p}^{*}\mathcal{M}$
so that we could obtain a 3-form:
\[
\mathbf{v}=\frac{1}{3}\left(\mathbf{l}_{1}\wedge\mathbf{l}_{2}\wedge\mathbf{l}_{3}\right),
\]
see FIG. 3. 
\begin{figure}[h]
\centering{}\includegraphics[scale=0.85]{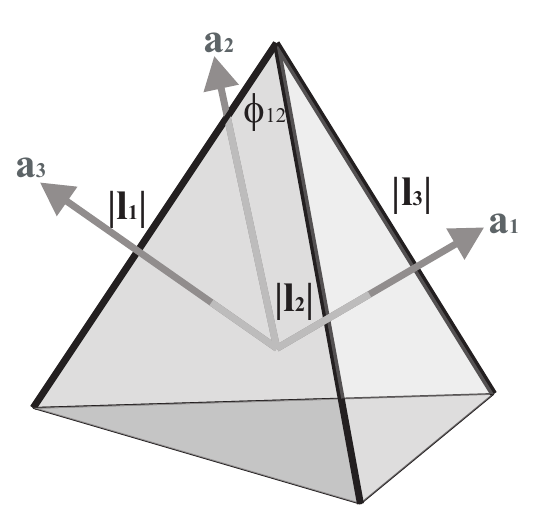}\caption[A tetrahedron can be obtained by wedge-producting three 1-forms.]{A tetrahedron $\mathbf{v}$ can be obtained by wedge-producting three
1-forms: $\mathbf{l}_{1}$, $\mathbf{l}_{2}$, and $\mathbf{l}_{3}$.
$\phi_{ij}$ is the angle between these 1-forms. The wedge product
between each two of these 1-forms, say, $\mathbf{l}_{i}$ and $\mathbf{l}_{j}$,
construct a triangle $\mathbf{a}_{k}.$ }
\end{figure}

The norm of $\mathbf{v}$ can be obtained by using the metric $g$
as in the triangle case:
\[
\left|\mathbf{v}\right|=\sqrt{g\left(\mathbf{v},\mathbf{v}\right)},
\]
and this can be interpreted as the \textit{volume of a tetrahedron}. 

Now we consider the subspaces of the tetrahedron. The first subspace
is the boundary $\partial\mathbf{v}$ which is the 2-dimensional \textit{space
of triangles}, consisting of four triangles:
\[
\partial\mathbf{v}=\left\{ \underset{\mathbf{a}_{1}}{\underbrace{\frac{1}{2}\left(\mathbf{l}_{2}\wedge\mathbf{l}_{3}\right)}},\underset{\mathbf{a}_{2}}{\underbrace{\frac{1}{2}\left(\mathbf{l}_{3}\wedge\mathbf{l}_{1}\right)}},\underset{\mathbf{a}_{3}}{\underbrace{\frac{1}{2}\left(\mathbf{l}_{1}\wedge\mathbf{l}_{2}\right)}},\underset{\mathbf{a}_{4}=-\mathbf{a}_{123}}{\underbrace{-\frac{1}{2}\left(\mathbf{l}_{1}\wedge\mathbf{l}_{2}+\mathbf{l}_{2}\wedge\mathbf{l}_{3}+\mathbf{l}_{3}\wedge\mathbf{l}_{1}\right)}}\right\} ,
\]
where the boundary satisfies \textit{closure condition}:
\begin{equation}
\sum_{i=1}^{4}\mathbf{a}_{i}=0,\label{eq:clos}
\end{equation}
which is also known as the Minkowski theorem \cite{key-3.5,key-3.26}.
This is a 2-form analogy to the closure condition in (\ref{eq:3.12}).
Another subspace of the tetrahedron is the \textit{space of segment},
which is 1-dimensional:
\[
\partial^{2}\mathbf{v}=\left\{ \mathbf{l}_{1},\mathbf{l}_{2},\mathbf{l}_{3},\mathbf{l}_{12},\mathbf{l}_{23},\mathbf{l}_{31}\right\} ,
\]
where the norms of every three segments $\left\{ \left|\mathbf{l}_{i}\right|,\left|\mathbf{l}_{j}\right|,\left|\mathbf{l}_{ij}\right|\right\} $
and $\left\{ \left|\mathbf{l}_{12}\right|,\left|\mathbf{l}_{23}\right|,\left|\mathbf{l}_{31}\right|\right\} $
satisfy triangle inequality (\ref{eq:triequal}), but only three triangles
satisfy the closure condition as in (\ref{eq:3.12}):
\begin{eqnarray*}
\mathbf{l}_{i}+\mathbf{l}_{j}-\mathbf{l}_{ij} & = & 0,\qquad i,j=1,2,3,\quad i\neq j,\\
\mathbf{l}_{12}+\mathbf{l}_{23}+\mathbf{l}_{31} & \neq & 0,\qquad\textrm{(in general).}
\end{eqnarray*}
The reason for this is because the boundary $\partial\mathbf{v}$
is a closed surface homeomorphic to a sphere $\mathcal{S}^{2},$ and
to cover a sphere, we need minimal two charts, while in this derivation
we only use a single vector space which is $T_{p}^{*}\mathcal{M};$
in other words, we need $\left\{ \mathbf{l}_{12},\mathbf{l}_{23},\mathbf{l}_{31}\right\} $
to live in another vector space $T_{q}^{*}\mathcal{M},$ if we want
to force them to satisfy triangle inequality.

It can be shown that the wedge product of two 1-forms (and their permutations)
meeting at a same point is a triangle: 
\begin{eqnarray*}
\frac{1}{2}\left(\mathbf{l}_{i}\wedge\mathbf{l}_{j}\right) & = & \mathbf{a}_{k},\qquad i,j,k=1,2,3,\quad i\neq j\neq k.
\end{eqnarray*}
There is a beautiful hierarchial structure among a simplex with its
subspaces.

\subsubsection{4-simplex}

A 4-simplex is a 4-dimensional analog of a triangle in 2-dimension
and a tetrahedron in 3-dimension. Its 2-dimension projection is illustrated
in FIG. 4(a). A 4-simplex can only be realized in space with dimension
$n\geq4.$ To build a 4-simplex $\mathbf{s}$, we need four distinct
1-forms, $\left\{ \mathbf{l}_{1},\mathbf{l}_{2},\mathbf{l}_{3},\mathbf{l}_{4}\right\} :$
\[
\mathbf{s}=\frac{1}{4}\left(\mathbf{l}_{1}\wedge\mathbf{l}_{2}\wedge\mathbf{l}_{3}\wedge\mathbf{l}_{4},\right),
\]
see FIG. 4(a). 
\begin{figure}[h]
\centering{}\includegraphics[scale=0.85]{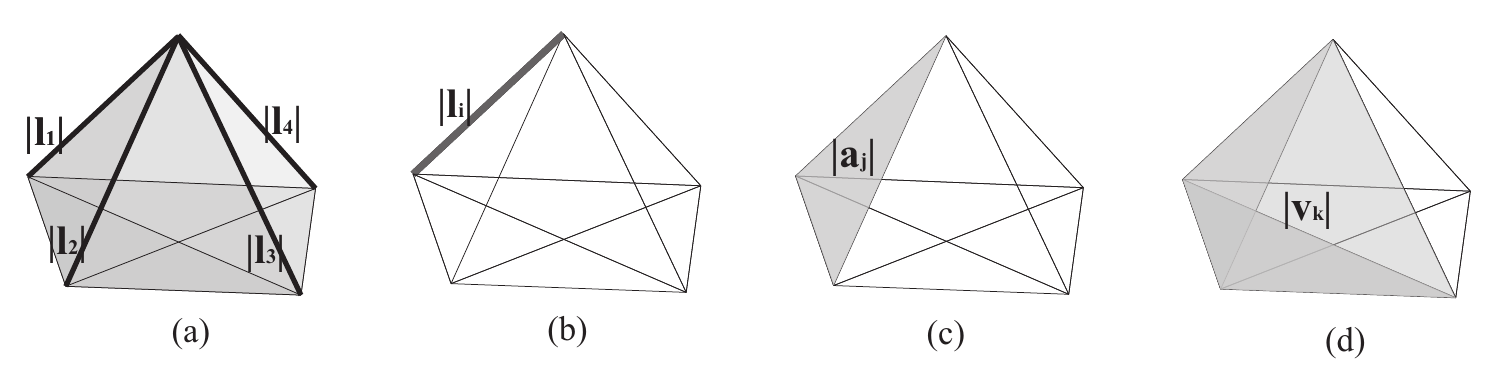}\caption[A 4-simplex can be obtained by wedge-producting four 1-forms]{(a) Construction of a 4-simplex by wedge-producting four 1-forms:
$\mathbf{l}_{1}$, $\mathbf{l}_{2}$, $\mathbf{l}_{3}$. and $\mathbf{l}_{4}$.
(b) the 1-dimensional subspace of a 4-simplex $\mathbf{s}$ consists
ten segments $\mathbf{l}_{i}$. (c) the 2-dimensional subspace of
$\mathbf{s}$ contains also ten triangle $\mathbf{a}_{j}$. (d) the
3-dimensional subspace of $\mathbf{s}$, which is also its boundary,
contains five tetrahedra $\mathbf{v}_{k}$. }
\end{figure}
 The norm of the 4-form $\mathbf{s}$ is:
\[
\left|\mathbf{s}\right|=\sqrt{g\left(\mathbf{s},\mathbf{s}\right)},
\]
and interpreted as the \textit{4-dimensional 'volume' of a 4-simplex.}

Similar with the previous cases, we could obtain subspaces of the 4-simplex,
which now has three types of subspaces. The first subspace is the
1-dimensional subspace, consisting of ten segments:
\[
\partial^{3}\mathbf{s}=\left\{ \mathbf{l}_{1},\mathbf{l}_{2},\mathbf{l}_{3},\mathbf{l}_{4},\mathbf{l}_{12},\mathbf{l}_{13},\mathbf{l}_{14},\mathbf{l}_{23},\mathbf{l}_{24},\mathbf{l}_{34}\right\} ,
\]
the norms of every three segments $\left\{ \left|\mathbf{l}_{i}\right|,\left|\mathbf{l}_{j}\right|,\left|\mathbf{l}_{ij}\right|\right\} $
and $\left\{ \left|\mathbf{l}_{ij}\right|,\left|\mathbf{l}_{jk}\right|,\left|\mathbf{l}_{ki}\right|\right\} $
satisfy triangle inequality (\ref{eq:triequal}), but only six from
ten triangles satisfy the closure condition as in (\ref{eq:3.12}):
\begin{eqnarray*}
\mathbf{l}_{i}+\mathbf{l}_{j}-\mathbf{l}_{ij} & = & 0,\qquad i,j=1,2,3,\quad i\neq j,\\
\mathbf{l}_{ij}+\mathbf{l}_{jk}+\mathbf{l}_{ki} & \neq & 0,\qquad\textrm{(in general).}
\end{eqnarray*}
The reason for this is the same with the reason in the lower dimensional
case in the previous section.

The second subspace is the 2- dimensional subspace, consisting ten
triangles:
\[
\partial^{2}\mathbf{s}=\left\{ \mathbf{a}_{12},\mathbf{a}_{13},\mathbf{a}_{14},\mathbf{a}_{23},\mathbf{a}_{24},\mathbf{a}_{34},\mathbf{a}_{123},\mathbf{a}_{124},\mathbf{a}_{134},\mathbf{a}_{234}\right\} ,
\]
where:
\begin{eqnarray*}
\mathbf{a}_{ij} & = & \frac{1}{2}\left(\mathbf{l}_{i}\wedge\mathbf{l}_{j}\right),\qquad i,j,k=1,2,3,4\quad i\neq j\neq k.\\
\mathbf{a}_{ijk} & = & -\frac{1}{2}\left(\mathbf{l}_{i}\wedge\mathbf{l}_{j}+\mathbf{l}_{j}\wedge\mathbf{l}_{k}+\mathbf{l}_{k}\wedge\mathbf{l}_{i}\right).
\end{eqnarray*}
Each four of the triangles construct a boundary of a tetrahedron:
\[
\partial\mathbf{v}_{l}=\left\{ \mathbf{a}_{ij},\mathbf{a}_{jk},\mathbf{a}_{ki},\mathbf{a}_{ijk}\right\} ,\qquad i,j,k,l=1,2,3,4\quad i\neq j\neq k\neq l,
\]
\textit{i.e}., $\left\{ \left|\mathbf{a}_{ij}\right|,\left|\mathbf{a}_{jk}\right|,\left|\mathbf{a}_{ki}\right|,\left|\mathbf{a}_{ijk}\right|\right\} $
and $\left\{ \left|\mathbf{a}_{123}\right|,\left|\mathbf{a}_{124}\right|,\left|\mathbf{a}_{134}\right|,\left|\mathbf{a}_{234}\right|\right\} $
satisfy triangle inequality for higher forms, but only $\left\{ \mathbf{a}_{ij},\mathbf{a}_{jk},\mathbf{a}_{ki},\mathbf{a}_{ijk}\right\} $
satisfy the closure condition for the same reason as the previous
cases.

The last one is the 3- dimensional subspace which is the boundary
of the 4-simplex, consisting five tetrahedra:
\[
\partial\mathbf{s}=\left\{ \mathbf{v}_{1},\mathbf{v}_{2},\mathbf{v}_{3},\mathbf{v}_{4},\mathbf{v}_{5}=-\mathbf{v}_{1234}\right\} ,
\]
with:
\begin{eqnarray*}
\mathbf{v}_{l} & = & \mathbf{l}_{i}\wedge\mathbf{l}_{j}\wedge\mathbf{l}_{k},\qquad i,j,k,l=1,2,3,4\quad i\neq j\neq k\neq l\\
\mathbf{v}_{1234} & = & -\left(\sum_{l=1}^{4}\mathbf{v}_{l}\right).
\end{eqnarray*}
The boundary satisfies 3-form closure condition:
\[
\sum_{l=1}^{5}\mathbf{v}_{l}=0.
\]

This definition can be easily generalized to $p$-simplex in any dimension.

\subsection{Dihedral angles}

In Regge geometry, where we have discrete manifold instead of continuous
manifold, the intrinsic curvature is defined by angles \cite{key-3.19,key-3.23}.
In this subsection, we will review the definition of angles on a simplices.

\subsubsection{The dihedral angles}

Having an inner product defined in the space of forms $\Omega^{p}\left(\mathbb{R}^{n}\right)$
using a Euclidean metric $g$, we could have a notion of \textit{spherical
angle}. In general, spherical angle is defined by relation:
\begin{equation}
\cos\phi_{ab}=\frac{g\left(\mathbf{a},\mathbf{b}\right)}{\left|\mathbf{a}\right|\left|\mathbf{b}\right|},\label{eq:3.17}
\end{equation}
given vectors $\mathbf{a},\mathbf{b},$ and a Riemannian metric $g.$
Since $\Omega^{p}\left(\mathbb{R}^{n}\right)$ is also a vector space,
we could use (\ref{eq:3.17}) to define angles in the space of forms
$\Omega^{p}\left(\mathbb{R}^{n}\right)$.

Let $\mathbf{s}\in\Omega^{4}\left(\mathbb{R}^{4}\right)$ be a 4-form
describing a 4-simplex. Inside $\mathbf{s}$, there are segments $\mathbf{l}_{i}$,
triangles $\mathbf{a}_{j}$, and tetrahedra $\mathbf{v}_{k}$ as a
subspaces of $\mathbf{s.}$ These are 1-forms, 2-forms, and 3-forms
respectively. Since in $\Omega^{4}\left(\mathbb{R}^{4}\right)$, $\mathbf{l}_{i}$,
$\mathbf{a}_{j}$, and $\mathbf{v}_{k}$ have directions (they have
components), they act as vectors, not as scalars. Therefore, we can
have three notions of angle inside a 4-simplex $\mathbf{s}$: angles
between segments, between triangles, and between tetrahedra. These
angles are always located around a \textit{hinges}, which are the
$\left(p-2\right)$-simplices \cite{key-3.19,key-3.23}.

\paragraph{Dihedral angle on a point. }

This angle is defined as the spherical angle between two segments.
Given 1-forms $\mathbf{l}_{i}$ and $\mathbf{l}_{j}$ $\in T_{p}^{*}\mathcal{M}$,
the \textit{2-dimensional dihedral angle }$\phi_{ij}$ on a point
is defined as:
\begin{equation}
\phi_{ij}=\pi-\bar{\phi}_{ij},\qquad\cos\bar{\phi}_{ij}=\frac{g\left(\mathbf{l}_{i},\mathbf{l}_{j}\right)}{\left|\mathbf{l}_{i}\right|\left|\mathbf{l}_{j}\right|}.\label{eq:3x}
\end{equation}
These angles are located around point $p$ of the 4-simplex, see FIG.
5(a).

\paragraph{Dihedral angle on a segment.}

Another angle we have in a 4-simplex is the angle between two triangles
(we usually called it as 'dihedral' angle). This angle is defined
by an intersection of two planes, meeting on a segment. Given 2-forms
$\mathbf{a}_{i},\mathbf{a}_{j}$, the \textit{3-dimensional dihedral
angle }$\theta_{ij}$ on a segment is defined as:
\begin{equation}
\theta_{ij}=\pi-\bar{\theta}_{ij},\qquad\cos\bar{\theta}_{ij}=\frac{g\left(\mathbf{a}_{i},\mathbf{a}_{j}\right)}{\left|\mathbf{a}_{i}\right|\left|\mathbf{a}_{j}\right|},\label{eq:3y}
\end{equation}
where the inner product of forms is defined in Subsection III A. See
FIG. 5(b). 
\begin{figure}[h]
\centering{}\includegraphics[scale=0.85]{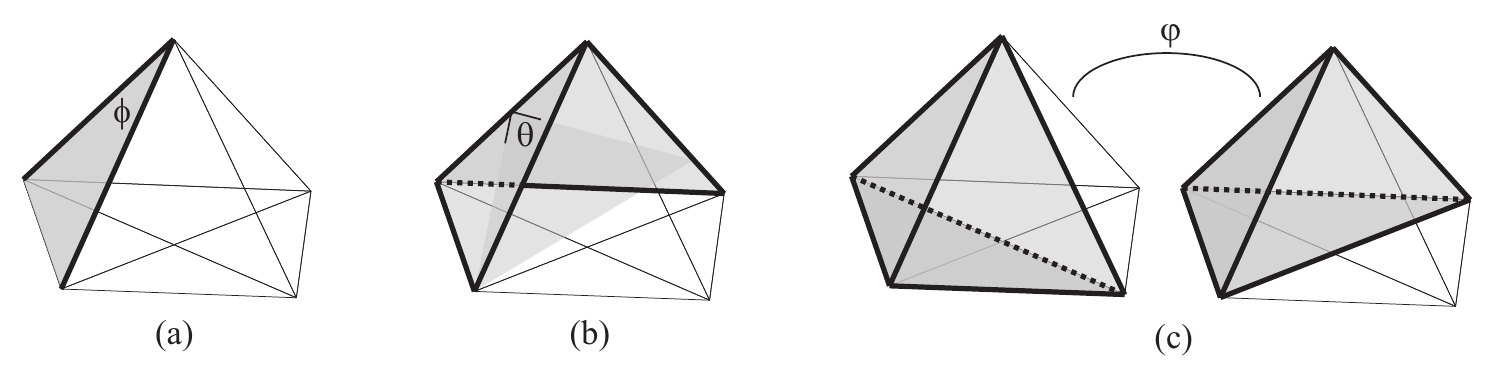}\caption[Types of dihedral angle inside a 4-simplex.]{Types of dihedral angle inside a 4-simplex. (a) $\phi$ is the 2-dimensional
dihedral angle\textit{ }between two segments, located on a point of
the 4-simplex. (b) $\theta$ is the 3-dimensional dihedral angle\textit{
}between two triangles, located on a segment of the 4-simplex. (c)
$\varphi$ is the 4-dimensional dihedral angle\textit{ }between two
tetrahedra, located on a triangle of the 4-simplex. }
\end{figure}

\paragraph{Dihedral angle on a plane. }

This angle is not common in standard 3-dimensional geometry. It comes
from an intersection between two tetrahedra, meeting on a plane. Remember
that in 4-dimension or larger, a 3-dimensional geometric figures defined
by 3-forms have directions, since the 3-form is not yet a volume form
in this space. On 4-dimension, these 3D geometrical figures live in
a 4-dimensional vector space spanned by basis $dx^{i}\wedge dx^{j}\wedge dx^{k}.$

Given 3-forms $\mathbf{v}_{i},\mathbf{v}_{j}$, the \textit{4-dimensional
dihedral angle }$\varphi_{ij}$ on a plane is defined as:
\[
\varphi_{ij}=\pi-\bar{\varphi}_{ij},\qquad\cos\bar{\varphi}_{ij}=\frac{g\left(\mathbf{v}_{i},\mathbf{v}_{j}\right)}{\left|\mathbf{v}_{i}\right|\left|\mathbf{v}_{j}\right|},
\]
see FIG. 5(c).

In 4-dimension, we can only have forms up to 4-form: $dx^{\mu}\wedge dx^{\nu}\wedge dx^{\sigma}\wedge dx^{\rho}$
, which is a volume-form in $\Omega^{4}\left(\mathbb{R}^{4}\right).$
No higher forms of geometry can be constructed. Therefore, we can
only have three types of dihedral angles which are: $\phi$, the angles
between segments meeting at a point; $\theta$, the angles between
planes meeting at a segments; and $\varphi,$ the angles between 3D
spaces meeting on a plane. The next step is to obtain the relation
between these dihedral angles through the 'dihedral angle formula'.

\subsubsection{Dihedral angle formula}

In the standard 3-dimensional Euclidean geometry, we have the remarkable
dihedral angle formula of the tetrahedron, which is a relation between
$\phi$, the angles between segments of the tetrahedron, and $\theta,$
the angles between planes of the same tetrahedron \cite{key-3.27}:
\begin{equation}
\cos\theta{}_{ij,k}=\frac{\cos\phi{}_{ij}-\cos\phi_{ik}\cos\phi_{kj}}{\sin\phi_{ik}\sin\phi_{kj}},\label{eq:8}
\end{equation}
see FIG. 6. 
\begin{figure}[h]
\centerline{\includegraphics[height=3.5cm]{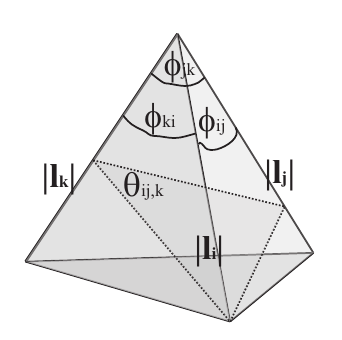}} \caption[Angles on tetrahedron.]{Given angles $\phi_{ij}$, $\phi_{ik}$, $\phi_{jk}$ at point $p$
of a tetrahedron, we could obtain the dihedral angle $\theta{}_{ij,k}$.
In fact, $\theta{}_{ij,k}$ is only $\phi_{ij}$ projected on the
plane normal to segment $\left|\mathbf{l}_{k}\right|$. }
\end{figure}
This relation can be derived algebraically using forms, in a relatively
simple way. See Appendix A.

Remarkably, as shown in \cite{key-3.28}, this dihedral angle relation
is valid for any dimension, which means it is a relation between a
$p$-dimensional dihedral angle (the angle between $\left(p-1\right)$-simplices)
with $\left(p-1\right)$-dimensional dihedral angle (the angle between
$\left(p-2\right)$-simplices). The dihedral angle formula can also
be written in the inverse form:
\begin{equation}
\cos\phi_{ij}=\frac{\cos\theta_{ij,k}-\cos\theta_{ik,j}\cos\theta_{kj,i}}{\sin\theta_{ik,j}\sin\theta_{kj,i}}.\label{eq:inverse}
\end{equation}

\section{Coordinate-free variables}

Segments, triangles, and tetrahedra, which are respectively, described
by 1-forms, 2-forms, and 3-forms, are basic geometrical elements in
3-dimensional discrete geometry. They are partitions of space: segments
are partitions of 1-dimensional space, triangles and more complex
polygons are partitions of 2-dimensional space, tetrahedra and polyhedra
are partitions of 3-dimensional space, and so on: $n$-polytopes are
partitions of $n$-dimensional space. In LQG, quantizing gravitational
field will give 'quanta of gravitational field', but since the gravitational
field is the spacetime itself; these 'quanta of gravitational field',
in the canonical framework, can be regarded as 'quanta' or 'particles'
of space. 

It is clear that in discrete geometry, we have a hierarchial structure
of spaces: \textit{A $p$-simplex is constructed from $\left(p+1\right)$
numbers of $\left(p-1\right)$-simplices}. We have already shown in
Subsection III A how a $p$-simplex can be constructed from several
numbers of lower-dimensional forms, by using the wedge product. This
is a \textit{vectorial} construction, where we explicitly use a specific
coordinate system. But since the theory of gravity (and all theory
of physics) needs to satisfy the \textit{general covariance} principle,
\textit{i.e.}, the formulation of any physical theory must be valid
in all coordinate system, it is convenient to propose a way to write
the set of 'coupled-particles' of spaces, as Regge had proposed, without
using coordinates at all: the \textit{coordinate-free variables}.
This will be the main task in this section.

\subsection{Single particle of space}

\subsubsection{The coordinate-free point of view}

A $p$-simplex living in an $n$-dimensional space can be described
using $p$ numbers of 1-forms, describing a $p$-form. The norm of
this $p$-form is interpreted as the $p$-dimensional \textit{volume}
of the $p$-simplex. See FIG. 7. 
\begin{figure}[h]
\centering{}\includegraphics[scale=0.85]{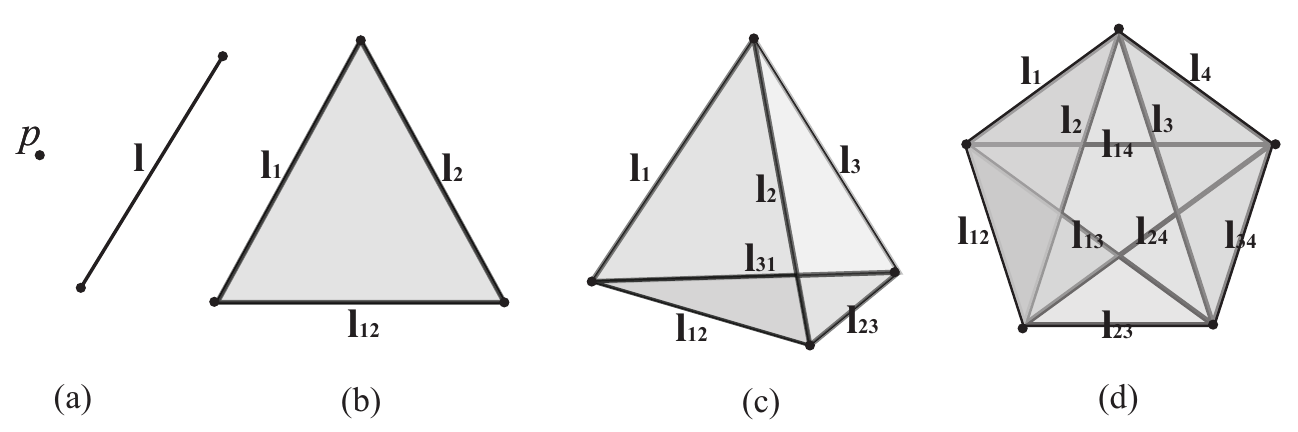}\caption[The vectorial picture to describe simplices.]{The vectorial picture to describe simplices: (a) A point is a $0$-simplex,
it is describe by a scalar and only has trivial information. A flat
line or a segment is a $1$-simplex, and it can be completely describe
by a 1-form $\mathbf{l}.$ (b) A triangle is a $2$-simplex, it is
completely determined by two 1-forms: $\mathbf{l}_{1}$ and $\mathbf{l}_{2}$,
the third 1-forms $\mathbf{l}_{12}$ comes form the addition of $\mathbf{l}_{1}$
and $\mathbf{l}_{2}$. (c) A tetrahedron is a 3-simplex, it is completely
determined by three 1-forms: $\mathbf{l}_{1},$ $\mathbf{l}_{2}$
and $\mathbf{l}_{3}$, the other three 1-forms, $\mathbf{l}_{ij},$
comes from the addition of $\mathbf{l}_{i}$ and $\mathbf{l}_{j}$.
(d) A 4-simplex is completely determined by four 1-forms, while the
other six 1-forms are the addition between two of these 1-forms. }
\end{figure}
 It is clear that a $p$-simplex living in $n$-dimensional space
needs $pn$ informations to describe these geometries in a vectorial
way, completely. These informations, depending on the $n$-dimensional
space where the $p$-form is embedded, are more than enough to describe
the geometries, because these informations also describe the position
and orientation of the geometries with respect to a specific origin
$\mathcal{O}$ of the vector space.

This vectorial picture to describe simplices (that is, by using forms)
as 'particles' of spaces, contradicts with one of our basic assumptions
used in general relativity: the \textit{background independence} \cite{key-3.34,key-3.35}.
The first contradiction is the forms live in an $n$-dimensional space
$\mathbb{R}^{n},$ which means if we describe the simplices using
these forms, then they are embedded in another 'backstage space' which
is $\mathbb{R}^{n}$. We do not want this since we want the simplices
to be particles of space. It \textit{creates} space and it should
be the space itself, without referring to any other background stage. 

Another contradiction is the notion of 'position' of the particles
of space. This is not satisfactory for the same reason: the particle
of space should be the space itself. It should not have a position
with respect to another background space, which in turns contradicts
the background independence. The 'location' of the particles of spaces
is defined in a 'relational' way, through the adjacency among particles. 

Let us find a way to write these geometrical objects in the coordinate-free
picture. See FIG. 8. 
\begin{figure}[h]
\centering{}\includegraphics[scale=0.85]{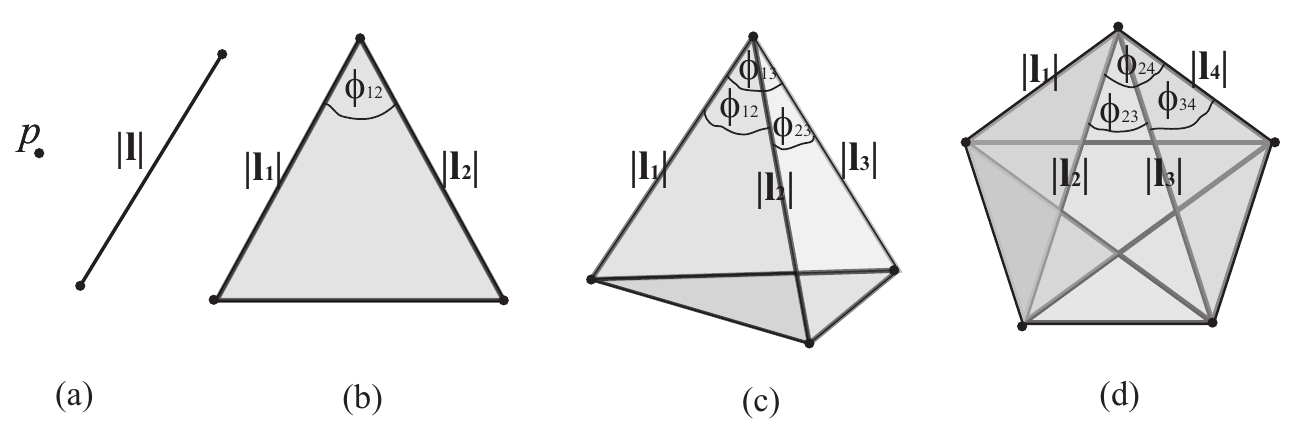}\caption[The coordinate-free picture to describe simplices.]{The coordinate-free picture to describe simplices. (a) A point is a $0$-simplex,
it is describe by a scalar, which is already coordinate independent.
(a) Instead of a 1-form, a flat line or a segment now is described
also by a scalar, which is the norm of the 1-form $\left|\mathbf{l}_{1}\right|$.
The segment is completely described by this single degrees of freedom,
without any information about the origin and the direction. (b) A
triangle, in a coordinate-free picture, can be completely determined by
three degrees of freedom/ informations: either the norms of the three
1-forms (which are the lengths of the segments of the triangle) $\left\{ \left|\mathbf{l}_{1}\right|,\left|\mathbf{l}_{2}\right|,\left|\mathbf{l}_{12}\right|\right\} $,
or two norms of 1-forms with the 2D dihedral angle between them: $\left\{ \left|\mathbf{l}_{1}\right|,\left|\mathbf{l}_{2}\right|,\phi_{12}\right\} $.
Knowing $\left|\mathbf{l}_{12}\right|$, we could obtain $\phi_{12}$
and vice versa. (c) A tetrahedron is completely determined by six
degrees of freedom/ informations: either the norms of the six 1-forms
$\left\{ \left|\mathbf{l}_{1}\right|,\right.$ $\left|\mathbf{l}_{2}\right|,$
$\left|\mathbf{l}_{3}\right|,$ $\left|\mathbf{l}_{12}\right|,$ $\left|\mathbf{l}_{23}\right|,$
$\left.\left|\mathbf{l}_{31}\right|\right\} $, or three norms of
1-forms with three 2D dihedral angles between two of them $\left\{ \left|\mathbf{l}_{1}\right|,\right.$
$\left|\mathbf{l}_{2}\right|,$ $\left|\mathbf{l}_{3}\right|,$ $\phi_{12},$
$\phi_{23},$$\left.\phi_{32}\right\} $. (d) At last, a 4-simplex
is completely determined by ten degrees of freedom/ informations:
either the norms of the ten 1-forms $\left\{ \left|\mathbf{l}_{i}\right|,\right.$
$\left|\mathbf{l}_{ij}\right|,$ $i,j=1,2,3,4,$ $\left.i\neq j\right\} $,
or four norms of 1-forms with six 2D dihedral angles between two of
them $\left\{ \left|\mathbf{l}_{i}\right|,\right.$ $\phi_{ij},$
$i,j=1,2,3,4,$ $\left.i\neq j\right\} $. }
\end{figure}
In this example, we use the lengths of the segments and the 2D dihedral
angles as coordinate-free variables. But to prevent ambiguities rising
for simplices larger than two, that is, to distinguish if our system
contains larger-dimensions simplices (triangles, tetrahedra, \textit{etc}..)
instead of only a set of segments, we use another variables which
give equivalent informations of the system. 

As an example, for a triangle, we use the following variables: $\left\{ \left|\mathbf{a}\right|,\left|\mathbf{l}_{1}\right|,\phi_{12}\right\} $
instead of $\left\{ \left|\mathbf{l}_{1}\right|,\left|\mathbf{l}_{2}\right|,\phi_{12}\right\} $,
where the area $\left|\mathbf{a}\right|$ will provide us information
that our system is a 2-simplex triangle, instead of a system consisting
two coupled segments (this will be explained in the next subsection).
$\left|\mathbf{a}\right|$ can be obtained by the following transformation:
\[
\left|\mathbf{a}\right|=\left|\mathbf{l}_{1}\right|\left|\mathbf{l}_{2}\right|\sin\phi_{12}.
\]
For the tetrahedron case, we can use the norm of a 3-form, which is
the volume of the tetrahedron, and the norms of the 2-forms, which
are areas of triangles, and the 3D dihedral angle between two triangles:
$\left\{ \left|\mathbf{v}\right|,\left|\mathbf{a}_{1}\right|,\left|\mathbf{a}_{2}\right|,\theta_{12},\theta_{23},\theta_{32}\right\} $
instead of lengths and 2D angles. The transformation is:
\[
\left|\mathbf{v}\right|=\frac{1}{3}\left|\mathbf{l}_{j}\right|\left|\mathbf{l}_{k}\right|\left|\mathbf{l}_{l}\right|\sqrt{1+2\cos\phi_{jk}\cos\phi_{kl}\cos\phi_{lj}-\left(\cos^{2}\phi_{jk}+\cos^{2}\phi_{kl}+\cos^{2}\phi_{lj}\right)},
\]
\begin{equation}
\left|\mathbf{a}_{i}\right|=\frac{1}{4}\sqrt{\left(\left|\mathbf{l}_{j}\right|+\left|\mathbf{l}_{k}\right|+\left|\mathbf{l}_{jk}\right|\right)\left(-\left|\mathbf{l}_{j}\right|+\left|\mathbf{l}_{k}\right|+\left|\mathbf{l}_{jk}\right|\right)\left(\left|\mathbf{l}_{j}\right|-\left|\mathbf{l}_{k}\right|+\left|\mathbf{l}_{jk}\right|\right)\left(\left|\mathbf{l}_{j}\right|+\left|\mathbf{l}_{k}\right|-\left|\mathbf{l}_{jk}\right|\right)},\label{eq:a}
\end{equation}
\begin{equation}
\cos\theta_{ij,k}=\frac{\cos\phi_{ij}-\cos\phi_{ik}\cos\phi_{kj}}{\sin\phi_{ik}\sin\phi_{kj}},\label{eq:dihed}
\end{equation}
with (\ref{eq:a}) is simply the Heron's formula for a triangle, while
(\ref{eq:dihed}) is simply the dihedral angle formula. There is a
lot of choice of variables to describe a tetrahedron, but we usually
choose the sets which gives a unique geometry.

For the 4-simplex case, we could use the norm of the 4-form, $\left|\mathbf{s}\right|,$
which is the volume of the 4-simplex, the volumes of tetrahedra, and
the 4D dihedral angle: $\left\{ \left|\mathbf{s}\right|,\right.$
$\left|\mathbf{v}_{i}\right|,$ $\left.\varphi_{ij}\right\} $, instead
of lengths and 2D angles (it could also be represented by using areas
and 3D angle just as in the previous 3D case). See FIG. 9. 
\begin{figure}[h]
\begin{centering}
\includegraphics[scale=0.85]{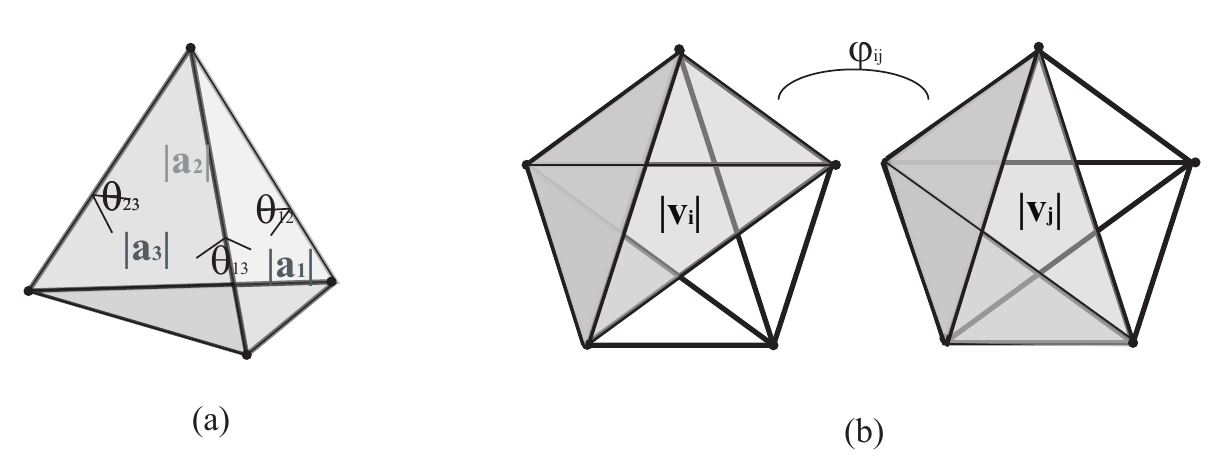}\caption[Coordinate-free variables for a single tetrahedron.]{(a) Instead of using lengths (of segments) and 2D angles (between
segments) $\left\{ \left|\mathbf{l}_{i}\right|,\phi_{ij}\right\} $,
we use the volume (of tetrahedron), areas (of the triangle) and 3D
angle (between triangles) $\left\{ \left|\mathbf{v}\right|,\left|\mathbf{a}_{i}\right|,\theta_{ij}\right\} $
as a coordinate-free variables for a single tetrahedron. (b) Adding
1-dimension higher, we use the \textit{volume form} of the 4-simplex,
3D volumes (of the tetrahedron) and 4D angle (between tetrahedra)
$\left\{ \left|\mathbf{s}\right|,\left|\mathbf{v}_{i}\right|,\varphi_{ij}\right\} $
as the coordinate-free variables for a single 4-simplex. }

\par\end{centering}

\end{figure}

From now on, we will use this coordinate-free picture to describe the
degrees of freedom of the geometries.

\subsubsection{The choice of variables and uniqueness of a simplex}

A simplex is uniquely determined by the lengths of its edges. It must
be kept in mind that if we wish to describe the simplices using other
variables different than their edges, these sets of variables need
to have a one-to-one map to the edges length.

In general, areas and volumes are polynomial (and nonlinear) functions
of the length of edges, so the map involving these variables to the
set of edges lengths of the simplex may be one-to-many, since polynomial
equations in general have more than one solution. This can be simply
illustrated in the 2-dimensional case of a triangle specified by its
area and two lengths, $\left\{ \left|\mathbf{a}\right|,\left|\mathbf{l}_{1}\right|,\left|\mathbf{l}_{2}\right|\right\} $.
This choice of variables does not uniquely describe a triangle: there
are two different triangles, specified with the three edge lengths
$\left\{ \left|\mathbf{l}_{1}\right|,\left|\mathbf{l}_{2}\right|,\left|\mathbf{l}_{12}\right|\right\} $
such that both have areas equal to $\mathbf{a}$. This is because
the equation expressing the third edge $\mathbf{l}_{12}$ in terms
of $\left\{ \left|\mathbf{a}\right|,\left|\mathbf{l}_{1}\right|,\left|\mathbf{l}_{2}\right|\right\} $
is a quadratic equation which have two solutions:
\[
\left|\mathbf{l}_{12}\right|^{2}=\left|\mathbf{l}_{1}\right|^{2}+\left|\mathbf{l}_{2}\right|^{2}-2\sqrt{\left|\mathbf{l}_{1}\right|^{2}\left|\mathbf{l}_{2}\right|^{2}-4\left|\mathbf{a}\right|^{2}},
\]
as long as the length is restricted to be definite positive.

We have classify all possible choices of variables which uniquely
describe a Euclidean triangle: $\left\{ \left|\mathbf{l}_{i}\right|,\phi_{ij},\phi_{ik}\right\} $
, $\left\{ \left|\mathbf{l}_{i}\right|,\phi_{ij},\phi_{jk}\right\} $,
$\left\{ \left|\mathbf{a}\right|,\phi_{ij},\phi_{ik}\right\} $, $\left\{ \left|\mathbf{l}_{i}\right|,\left|\mathbf{l}_{j}\right|,\phi_{ij}\right\} $,
and $\left\{ \left|\mathbf{a}\right|,\left|\mathbf{l}_{i}\right|,\phi_{ij}\right\} $.
Other choice of variables are not well-defined in the sense that the
information they contain do not describe uniquely a triangle.

Similar attempts could be done for tetrahedron and 4-simplex case.
For a tetrahedron, some unique and well-defined choice of variables
are: the volume, two areas of triangles, and three 3D dihedral angles
$\left\{ \left|\mathbf{v}\right|,\left|\mathbf{a}_{i}\right|,\left|\mathbf{a}_{j}\right|,\theta_{ij},\theta_{ik},\theta_{jk}\right\} $;
three areas and three 3D dihedral angles $\left\{ \left|\mathbf{a}_{i}\right|,\left|\mathbf{a}_{j}\right|,\left|\mathbf{a}_{k}\right|,\theta_{ij},\theta_{ik},\theta_{jk}\right\} $;
and four areas with two 3D dihedral angles $\left\{ \left|\mathbf{a}_{i}\right|,\left|\mathbf{a}_{j}\right|,\left|\mathbf{a}_{k}\right|,\left|\mathbf{a}_{ijk}\right|,\theta_{ij},\theta_{jk}\right\} $.
The case of a 4-simplex is already studied in \cite{key-3.19a},
where it turns out that the ten areas of triangles inside a 4-simplex
do not completely fix its geometry. We found that some of the unique
and well-defined choice of variables for a 4-simplex are: the 4-volume
of the 4-simplex, three 3-volumes of tetrahedra, and six 4D dihedral
angles $\left\{ \left|\mathbf{s}\right|,\left|\mathbf{v}_{i}\right|,\varphi_{jk}\right\} $,
$i=1,2,3$, $j,k=1,..,4,$ $j<k$; and four 3-volumes of tetrahedra
with six 4D dihedral angles $\left\{ \left|\mathbf{v}_{i}\right|,\varphi_{ij}\right\} $,
$i,j=1,..,4$, $i<j$. The proofs are given in the Appendix B.

\subsection{System of $n$-particles of space}

In this subsection, we will return back to the lengths of segments
and 2D angles variables, neglecting the ambiguity they brought, for
a reason which will be clear later. The use of these variables will
not rise any problem to our system, except the ambiguity for the set
of $p$-simplices, with $p\geq1$.

\subsubsection{Uncoupled system}

Degrees of freedom of a system containing uncoupled $n$-particles
of space can be obtained easily, by taking the degrees of freedom
of the single particle of space \textit{times} the number of the particles
in the system. For example, using coordinate-free variables, a single
segment is completely determined by a single degrees of freedom $\left|\mathbf{l}\right|$,
therefore, a system containing \textit{two} uncoupled segment, say,
segment $a$ and $b$, will contains \textit{two} degrees of freedom
$\left\{ \left|\mathbf{l}_{a}\right|,\left|\mathbf{l}_{b}\right|\right\} .$
For a higher dimensional case, a system containing three uncoupled
triangles $a$, $b$, $c,$ contains $3\times3=9$ degrees of freedom
which are $\left\{ \left|\mathbf{l}_{1}^{a}\right|,\left|\mathbf{l}_{2}^{a}\right|,\phi_{12}^{a},\left|\mathbf{l}_{1}^{b}\right|,\left|\mathbf{l}_{2}^{b}\right|,\phi_{12}^{b},\left|\mathbf{l}_{1}^{c}\right|,\left|\mathbf{l}_{2}^{c}\right|,\phi_{12}^{c}\right\} ;$
and a system containing four uncoupled tetrahedra $a$, $b$, $c,$
$d,$ contains $4\times6=24$ degrees of freedom, which are $\left\{ \left|\mathbf{l}_{1}^{i}\right|,\right.$
$\left|\mathbf{l}_{2}^{i}\right|,$ $\left|\mathbf{l}_{3}^{i}\right|,$
$\phi_{12}^{i},$ $\phi_{23}^{i},$ $\left.\phi_{32}^{i}\right\} $
for $i=a,b,c,d.$ See FIG. 10. 
\begin{figure}[h]
\begin{centering}
\includegraphics[scale=0.85]{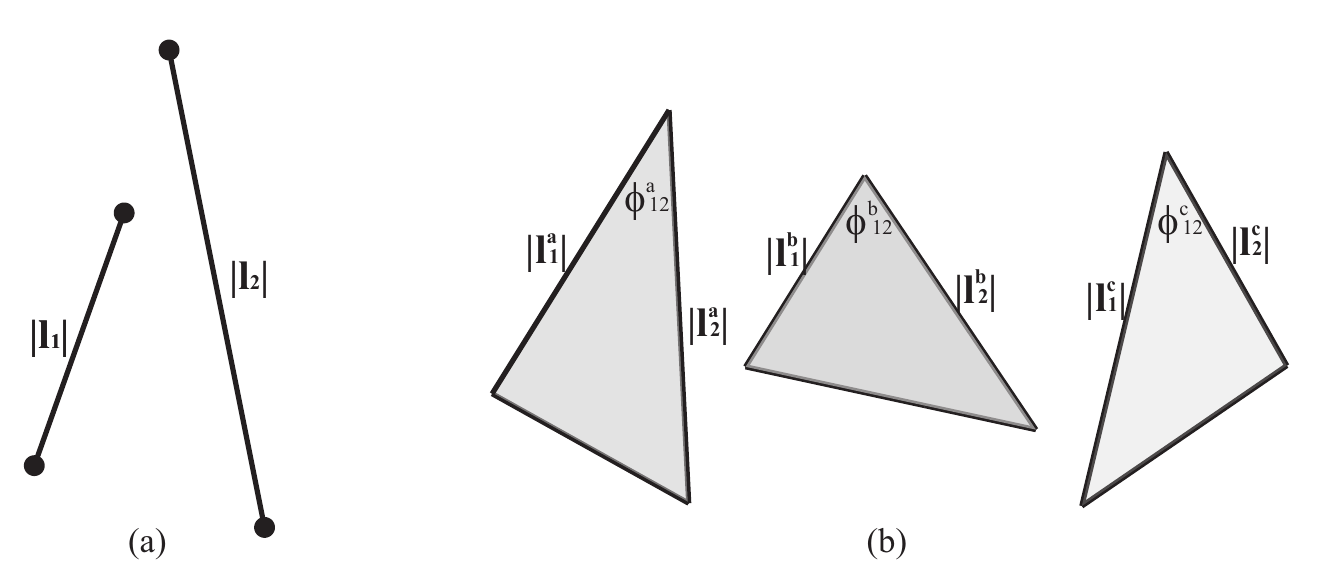}\caption[(a) A system of two uncoupled segments, and (b) a system of three
uncoupled triangles.]{(a) A system of two uncoupled segments $a$ and $b$, containing two
degrees of freedom $\left\{ \left|\mathbf{l}_{a}\right|,\left|\mathbf{l}_{b}\right|\right\} .$
(b) A system of three uncoupled triangles $a$, $b$, $c,$ containing
nine degrees of freedom which are $\left\{ \left|\mathbf{l}_{1}^{a}\right|,\left|\mathbf{l}_{2}^{a}\right|,\phi_{12}^{a},\left|\mathbf{l}_{1}^{b}\right|,\left|\mathbf{l}_{2}^{b}\right|,\phi_{12}^{b},\left|\mathbf{l}_{1}^{c}\right|,\left|\mathbf{l}_{2}^{c}\right|,\phi_{12}^{c}\right\} .$
}

\par\end{centering}

\end{figure}

\subsubsection{Coupled system}

We consider \textit{coupled} system. To couple two degrees of freedom,
we need a '\textit{coupling}', moreover, we define the coupling as
a dynamical variable, where each coupling will add another single
degrees of freedom to the system. Suppose we have
two particles with degrees of freedom $x_{a}$ and $x_{b}$. The natural
coupling would be in terms of $g_{ab}$, with $0\leq\left|g_{ab}\right|\leq1$.
$0$ means 'no interaction' between particle $a$ and $b$, while
$1$ means 'maximal interaction' between them. 

In a system of particle of space, we define the \textit{coupling $g$
}to be the \textit{$(p+1)$-dimensional angle between two $p$-simplices},
\textit{located on the $(p-1)$-simplex}. Let us take an example;
a system of two coupled segments, say, segment $a$ and $b$, will
contain three degrees of freedom $\left\{ \left|\mathbf{l}_{a}\right|,\left|\mathbf{l}_{b}\right|,g_{ab}=\cos\phi_{ab}\right\} ,$
together with the coupling constant. For a higher dimensional case,
a system of three coupled triangles $a$, $b$, $c,$ contains twelve
degrees of freedom which are $\left\{ \left|\mathbf{l}_{1}^{a}\right|,\left|\mathbf{l}_{2}^{a}\right|,\phi_{12}^{a},\left|\mathbf{l}_{1}^{b}\right|,\right.$$\left.\left|\mathbf{l}_{2}^{b}\right|,\phi_{12}^{b},\left|\mathbf{l}_{1}^{c}\right|,\left|\mathbf{l}_{2}^{c}\right|,\phi_{12}^{c}\right\} $,
together with the coupling constant $\left\{ g_{ab}=\cos\theta_{ab}\right.,$
$g_{ac}=\cos\theta_{ac},$ $\left.g_{bc}=\cos\theta_{bc}\right\} .$
At last, a system containing four coupled tetrahedra $a$, $b$, $c,$
$d,$ contains thirty degrees of freedom which are $\left\{ \left|\mathbf{l}_{1}^{i}\right|,\left|\mathbf{l}_{2}^{i}\right|,\left|\mathbf{l}_{3}^{i}\right|,\phi_{12}^{i},\phi_{23}^{i},\phi_{32}^{i}\right\} ,$
$i=a,b,c,d,$ together with the coupling constant $\left\{ g_{ij}=\cos\varphi_{ij}\right\} ,$
with $i,j=a,b,c,d,$ $i\neq j$. See FIG. 11. 
\begin{figure}[h]
\centering{}\includegraphics[scale=0.85]{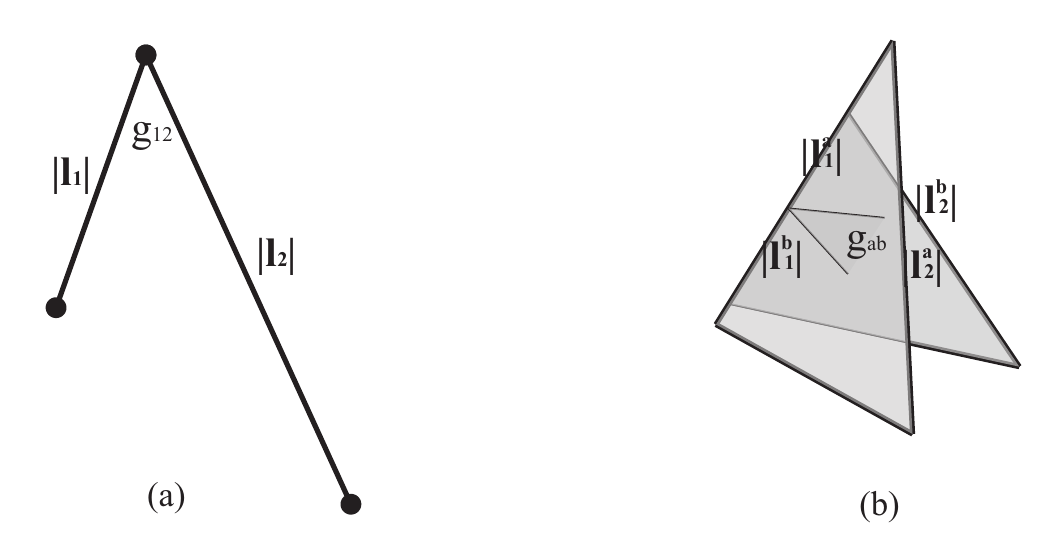}\caption[(a) A system of two coupled segments, and (b) a system of two coupled
triangles.]{(a) A system containing two coupled segments $a$ and $b$, containing
three degrees of freedom $\left\{ \left|\mathbf{l}_{a}\right|,\left|\mathbf{l}_{b}\right|,g_{ab}=\cos\phi_{ab}\right\} ,$
with the coupling constant $g_{ab}$. (b) A system containing two
coupled triangles $a$ and $b$, containing seven degrees of freedom
which are $\left\{ \left|\mathbf{l}_{1}^{a}\right|,\right.$ $\left|\mathbf{l}_{2}^{a}\right|,$
$\phi_{12}^{a},$ $\left|\mathbf{l}_{1}^{b}\right|,$ $\left|\mathbf{l}_{2}^{b}\right|,$
$\phi_{12}^{b},$ $\left.g_{ab}=\cos\theta_{ab}\right\} ,$ with the
coupling constant $g_{ab}$. }
\end{figure}

\subsubsection{Constraint: shape-matching condition}

\textit{Constraints} are specific conditions that must be satisfied
by a system or a part of a system. Imposing constraints will \textit{reduce}
the degrees of freedom in a system as many as the number of constraints
added \cite{key-3.36,key-3.37}. In this subsection, we will impose
a constraint known as the \textit{shape-matching constraint} \cite{key-3.5,key-3.28},
which guarantees the shape of the $\left(p-1\right)$-simplex where
the two $p$-simplex meet to be \textit{exactly} the same. A $\left(p-1\right)$-simplex
contains $\frac{p\left(p-1\right)}{2}$ segments (which are $1$-simplices),
and two $\left(p-1\right)$-simplices (by neglecting their reflection
symmetries) are exactly the same \textit{iff} all the norm of their
$\frac{p\left(p-1\right)}{2}$ segments (which are $1$-simplices)
are the same:
\begin{equation}
\left|\mathbf{l}_{i}^{a}\right|\equiv\left|\mathbf{l}_{i}^{b}\right|,\qquad i=1,...,\frac{p\left(p-1\right)}{2}.\label{eq:shape}
\end{equation}
Therefore, giving a shape-matching constraint to two coupled $p$-simplices
meeting on a common $\left(p-1\right)$-simplex, will reduce their
degrees of freedom as many as $\frac{p\left(p-1\right)}{2}$ degrees
of freedom. See FIG. 12 as an example. 
\begin{figure}[h]
\begin{centering}
\includegraphics[scale=0.85]{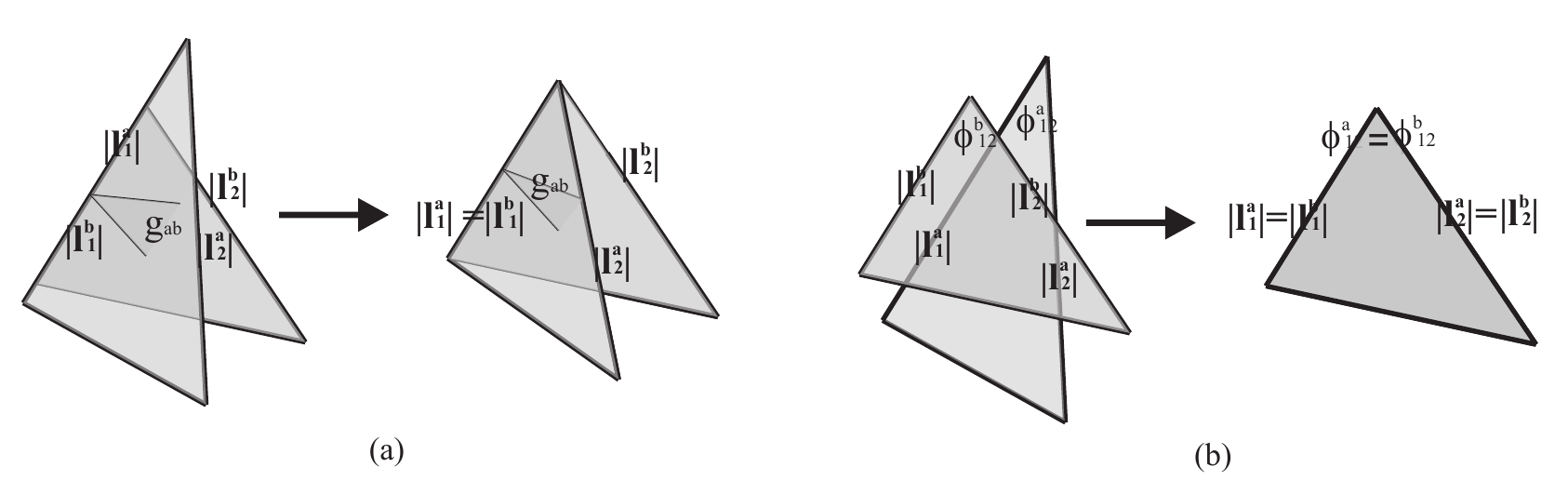}
\par\end{centering}

\caption[Examples of constraint conditions on set of simplices.]{(a) A system of two coupled triangles $a$ and $b$, containing seven
degrees of freedom which are $\left\{ \left|\mathbf{l}_{1}^{a}\right|,\right.$
$\left|\mathbf{l}_{2}^{a}\right|,$ $\phi_{12}^{a},$ $\left|\mathbf{l}_{1}^{b}\right|,$
$\left|\mathbf{l}_{2}^{b}\right|,$ $\phi_{12}^{b},$ $\left.g_{ab}=\cos\theta_{ab}\right\} ,$
before imposing constraint on the segment. After imposing the constraint,
$\left|\mathbf{l}_{1}^{a}\right|\equiv\left|\mathbf{l}_{1}^{b}\right|,$
the degrees of freedom reduce from seven to six. (b) A system containing
two coupled tetrahedra $a$ and $b$, containing thirteen degrees
of freedom which are $\left\{ \left|\mathbf{l}_{1}^{i}\right|,\right.$
$\left|\mathbf{l}_{2}^{i}\right|,$ $\left|\mathbf{l}_{3}^{i}\right|,$
$\phi_{12}^{i},$ $\phi_{23}^{i},$ $\phi_{32}^{i},$ $\left.g_{ab}=\cos\varphi_{ab}\right\} ,$
for $i=a,b,$ before imposing constraint on the segments. After imposing
three constraints: $\left|\mathbf{l}_{k}^{a}\right|\equiv\left|\mathbf{l}_{k}^{b}\right|,$
$k=1,2,$ and $\phi_{12}^{a}=\phi_{12}^{b}$ on the triangle where
the two tetrahedra meet, the degrees of freedom reduce from thirteen
to ten. }

\end{figure}

For a system of simplices containing number of particles $n\geq3$,
the shape matching condition does not satisfy (\ref{eq:shape}), they
are much more complicated because we need to be careful not to overcount
a same constraint equation twice. The constraint could contain also
the dihedral angle relation, restricting the choice of the coupling
constant for not being arbitrary, but satisfying (\ref{eq:dihed}).
Generally, the number of constraints will depend on \textit{how} each
simplex couples to each other. The next task is to obtain the formula
for calculating the degrees of freedom.

\subsection{Calculating the degrees of freedom}

In general, the total degrees of freedom of a system of $n$-particles
can be obtained by the following formula:
\[
N_{\textrm{d.o.f.}}=nN_{\textrm{degeneracy}}+N_{\textrm{coupling}}-N_{\textrm{constraint}},
\]
with $N_{\textrm{degeneracy}}$ is the number of degeneracy inside
a single particle, $N_{\textrm{coupling}}$ is the number of the coupling
constant, and $N_{\textrm{constraint}}$ is the number of the constraints. 

For a system of $n$-particle of space, $N_{\textrm{degeneracy}}$
will depend on the simplices we use; for example, if it is a system
of $1$-simplices, a $1$-simplex have \textit{no} degeneracy, because
a segment has only single degrees of freedom. A $2$-simplex have
$N_{\textrm{degeneracy}}=3,$ since a triangle have three degrees
of freedom. A $3$-simplex have $N_{\textrm{degeneracy}}=6,$ since
a tetrahedron have six degrees of freedom. Generalizing, a $p$-simplex
have $\frac{p\left(p+1\right)}{2}$ degrees of freedom and therefore:
\[
N_{\textrm{degeneracy}}=\frac{p\left(p+1\right)}{2},
\]
for a $p$-simplex.

The number of coupling $N_{\textrm{coupling}}$ and the number of
the constraints $N_{\textrm{constraint}}$ of a system of $n$-particle
of space depend on how each simplex coupled to each other. In other
words, it depends on the \textit{configuration of the set} of simplices
to construct the portion of discrete space: it depends on the triangulation,
or more general, the \textit{tesselation} of the simplicial complex
\cite{key-3.38,key-3.39}. Because of this reason, we take a specific,
but very useful examples of configuration of simplices called as \textit{Pachner-moves},
see FIG. 13. 
\begin{figure}[h]

\begin{centering}
\includegraphics[scale=0.85]{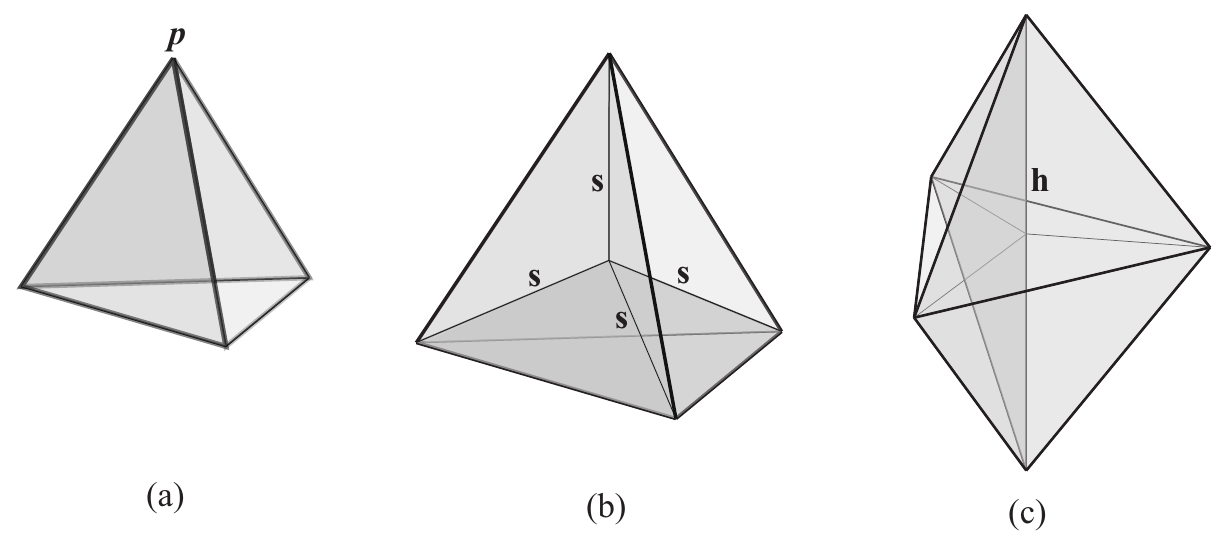}\caption[Examples of Pachner moves.]{(a) 3-1 Pachner move: three triangles are connected to each others
on their edges, constructing an open surface of a tetrahedron, without
the bottom triangle closing the surface. The 2D intrinsic curvature
is located at point $p$. (b) A 1-dimensional higher analog of (a),
the 4-1 Pachner move: four tetrahedra are connected to each others
on their internal triangles, constructing an open hypersurface of
a 4-simplex, without one tetrahedron closing the hypersurface. The
3D intrinsic curvatures are located on their four hinges: segments
$\mathbf{s}$. (c) 3-2 Pachner move: three tetrahedra are connected
to each others on their internal triangles, constructing a \textit{'trihedral}-\textit{bypiramid'}.
The 3D intrinsic curvature is located on segment $\mathbf{h}$. }

\par\end{centering}

\end{figure}

We calculate the degrees of freedom of these several Pachner-moves
cofigurations. See TABLE 1. 
\begin{table}[h]
\centering{}{\small{}}%
\begin{tabular}{|c|c|c|c|c|c|c|c|c|}
\hline 
{\small{$p$}} & {\small{$N_{\textrm{deg}}$ }} & {\small{moves}} & {\small{$n$}} & {\small{$N_{\textrm{d.o.f.}}$}} & {\small{$N_{\textrm{coupling}}$ }} & {\small{$N_{\textrm{d.o.f.}}$ }} & {\small{$N_{\textrm{const}}$}} & {\small{$N_{\textrm{d.o.f.}}$}}\tabularnewline
 &  &  &  & {\small{(uncoupled)}} &  & {\small{(coupled)}} &  & \tabularnewline
\hline 
\hline 
{\small{1 }} & {\small{1}} & {\small{2-1}} & {\small{2}} & {\small{$2\times1=2$}} & {\small{1}} & {\small{$2+1=3$}} & {\small{0}} & {\small{$3$}}\tabularnewline
\hline 
{\small{2 }} & {\small{3}} & {\small{3-1}} & {\small{3}} & {\small{$3\times3=9$}} & {\small{3}} & {\small{$9+3=12$}} & {\small{$3+3=6$}} & {\small{$12-6=6$}}\tabularnewline
\hline 
{\small{3 }} & {\small{6}} & {\small{4-1}} & {\small{4}} & {\small{$4\times6=24$}} & {\small{6}} & {\small{$24+6=30$}} & {\small{$14+6=20$}} & {\small{$30-20=10$}}\tabularnewline
\hline 
 &  & {\small{3-2}} & {\small{3}} & {\small{$3\times6=18$}} & {\small{3}} & {\small{$18+3=21$}} & {\small{$8+3=11$}} & {\small{$21-11=10$}}\tabularnewline
\hline 
\end{tabular}\caption{Degrees of freedom of several Pachner moves configurations.}
\end{table}
 For the 2-1 move, it is obvious that we have only one coupling $\phi$
and no constraint, so the total degrees of freedom is three, described by $\left\{ \left|\mathbf{l}_{1}\right|,\left|\mathbf{l}_{2}\right|,\phi_{12}\right\} .$
For the 3-1 move, we have three triangles where each of them has three
degrees of freedom; the number of coupling are three by using the
binomial relation $\binom{3}{2}$, the number of constraint are six:
three from relation (\ref{eq:shape}) for the segments where these
triangles meet, another three from the dihedral angle relation (\ref{eq:dihed})
on the coupling constants. Therefore the total degrees of freedom for 3-1
move is six, described by $\left\{ \left|\mathbf{a}_{1}\right|,\left|\mathbf{a}_{2}\right|,\left|\mathbf{a}_{3}\right|,\theta_{12},\theta_{13},\theta_{23}\right\} ,$
to remove ambiguity.

For the 4-1 move, we have four tetrahedron where each of them have
six degrees of freedom; the number of coupling are six by using the
binomial relation $\binom{4}{2}$, the number of constraint are twenty:
fourteen from relation (\ref{eq:shape}) for the segments, six from
the dihedral angle relation (\ref{eq:dihed}) for the coupling constants.
Therefore the total degrees of freedom for 4-1 move is ten, described by $\left\{ \left|\mathbf{v}_{i}\right|,\varphi_{ij}\right\} $,
$i\neq j=1,2,3,4$ . 

The last one, is the 3-2 move: we have three tetrahedron where each
of them have six degrees of freedom; the number of coupling are three,
the number of constraint are eleven: eight from relation (\ref{eq:shape})
for the segments, three from the dihedral angle relation (\ref{eq:dihed})
for the coupling constants. The total degrees of freedom for 3-2 move
is also ten, described by its three volumes of tetrahedra, 4D angles between them, three  edges length meeting on a common vertex, and a common segment of these tetrahedra: $\left\{ \left|\mathbf{v}_{i}\right|,\varphi_{ij},\left|\mathbf{l}_{i}\right|,\left|\mathbf{h}\right|\right\} ,$
$i\neq j=1,2,3.$ These variables can be proven to describe uniquely the geometry of the polyhedron.

\subparagraph*{Intrinsic curvatures as an emergent property.}

Given the definition of intrinsic curvature of discrete geometry,
which is the \textit{deficit angle} located on the hinge shared by
several simplices \cite{key-3.19,key-3.28,key-3.29}, it is clear
that curvature can only be defined in a system of coupled $n$-particles
of space, in other words, we could think the\textit{ intrinsic curvature
as an emergent property of a many-body system, it is the measure of
how strong is the 'interaction' among the 'particles' of space.}

\section{Conclusions}

Let us review our relatively new result obtained in this work: We
have obtain (1) a way of decribing a set of simplices vectorially by using differential forms,
(2) another way of describing a set of simplices using a coordinate-free picture, and (3) a consistent procedure to couple particles of space, together with a way to calculate the degrees
of freedom of the system of 'quanta' of space in the classical framework.

Our last result will be useful when we consider its application to
a coarse-graining method of discrete geometry. As a further work,
it is interesting if we could obtain a consistent procedure to couple
particles of space, and a way to calculate the degrees of freedom
of the system of 'quanta' of space in the twisted geometry framework.

\section*{Acknowledgment}
We thank Marko Vojinovic for his correction and suggestions concerning the uniqueness of the variables of simplices in Subsection III A.

\appendix

\section{(2+1)-dimensional dihedral angle relation}

Let $\mathbf{l}_{i},\mathbf{l}_{k}$ be vectors, instead of 1-forms,
that is, an element of $T_{p}\mathcal{M}$ instead of $T_{p}^{*}\mathcal{M}$
as in the previous derivation. Therefore we have two relations concerning
the dihedral angles:

\begin{equation}
g\left(\mathbf{l}_{i},\mathbf{l}_{k}\right)=\left|\mathbf{l}_{i}\right|\left|\mathbf{l}_{k}\right|\cos\phi_{ik},\label{eq:1}
\end{equation}
\begin{equation}
g\left(\mathbf{l}_{i}\wedge\mathbf{l}_{k},\mathbf{l}_{j}\wedge\mathbf{l}_{k}\right)=\left|\mathbf{l}_{i}\wedge\mathbf{l}_{k}\right|\left|\mathbf{l}_{j}\wedge\mathbf{l}_{k}\right|\cos\theta_{ij,k},\label{eq:2}
\end{equation}
with $\mathbf{l}_{i},$ $\mathbf{l}_{j},$ $\mathbf{l}_{k}$ are vectors.

\subparagraph*{Step 1: }

Using relation (\ref{eq:3.15}) concerning the components of a vector
2-forms, we obtain:

\begin{equation}
g\left(\mathbf{l}_{i}\wedge\mathbf{l}_{k},\mathbf{l}_{j}\wedge\mathbf{l}_{k}\right)=2\left(\left(l_{i}\right)_{\mu}\left(l_{k}\right)_{\nu}-\left(l_{i}\right)_{\nu}\left(l_{k}\right)_{\mu}\right)\left(\left(l_{j}\right)_{\rho}\left(l_{k}\right)_{\lambda}-\left(l_{j}\right)_{\lambda}\left(l_{k}\right)_{\rho}\right)\varepsilon^{\mu\nu}\varepsilon^{\rho\lambda},\label{eq:4}
\end{equation}
the basis $dx^{i}\wedge dx^{j}$ can be written as the Levi-Civita
symbol $\varepsilon^{ij}.$ In a flat space we can define the metric
as:
\[
g\left(dx^{\mu}\wedge dx^{\nu},dx^{\rho}\wedge dx^{\lambda}\right)=\varepsilon^{\mu\nu}\varepsilon^{\rho\lambda},
\]
following (\ref{eq:3aa}). Using the product of two Levi-Civita symbols
in (\ref{eq:3aaa}), we could write the right hand side of (\ref{eq:4})
as:
\[
2\left(\left(l_{i}\right)_{\mu}\left(l_{k}\right)_{\nu}-\left(l_{i}\right)_{\nu}\left(l_{k}\right)_{\mu}\right)\left(\left(l_{j}\right)_{\rho}\left(l_{k}\right)_{\lambda}-\left(l_{j}\right)_{\lambda}\left(l_{k}\right)_{\rho}\right)\left(\delta^{\mu\rho}\delta^{\nu\lambda}-\delta^{\mu\lambda}\delta^{\nu\rho}\right).
\]
Doing the tensor algebra, we obtain:
\begin{eqnarray}
g\left(\mathbf{l}_{i}\wedge\mathbf{l}_{k},\mathbf{l}_{j}\wedge\mathbf{l}_{k}\right) & = & \left(l_{i}\right)_{\mu}\left(l_{j}\right)^{\mu}\left(l_{k}\right)_{\nu}\left(l_{k}\right)^{\nu}-\left(l_{i}\right)_{\mu}\left(l_{k}\right)^{\mu}\left(l_{k}\right)_{\nu}\left(l_{j}\right)^{\nu},\nonumber \\
 & = & g\left(\mathbf{l}_{i},\mathbf{l}_{j}\right)g\left(\mathbf{l}_{k},\mathbf{l}_{k}\right)-g\left(\mathbf{l}_{i},\mathbf{l}_{k}\right)g\left(\mathbf{l}_{k},\mathbf{l}_{j}\right).\label{eq:5}
\end{eqnarray}
Using (\ref{eq:1}), we could write (\ref{eq:5}) as:
\begin{equation}
g\left(\mathbf{l}_{i}\wedge\mathbf{l}_{k},\mathbf{l}_{j}\wedge\mathbf{l}_{k}\right)=\left|\mathbf{l}_{i}\right|\left|\mathbf{l}_{k}\right|^{2}\left|\mathbf{l}_{j}\right|\left(\cos\phi_{ij}-\cos\phi_{ik}\cos\phi_{kj}\right).\label{eq:6}
\end{equation}

\subparagraph{Step 2:}

Now, we try to derive the expression for $\left|\mathbf{l}_{i}\wedge\mathbf{l}_{k}\right|.$
We know that: 
\[
g\left(\mathbf{l}_{i}\wedge\mathbf{l}_{k},\mathbf{l}_{j}\wedge\mathbf{l}_{k}\right) =\left|\mathbf{l}_{i}\wedge\mathbf{l}_{k}\right|^{2},
\]
so using (\ref{eq:5}):

\[
\left|\mathbf{l}_{i}\wedge\mathbf{l}_{k}\right|^{2}=g\left(\mathbf{l}_{i},\mathbf{l}_{i}\right)g\left(\mathbf{l}_{k},\mathbf{l}_{k}\right)-g\left(\mathbf{l}_{i},\mathbf{l}_{k}\right)g\left(\mathbf{l}_{k},\mathbf{l}_{i}\right),
\]
then using (\ref{eq:1}):
\[
\left|\mathbf{l}_{i}\wedge\mathbf{l}_{k}\right|^{2}=\left|\mathbf{l}_{i}\right|^{2}\left|\mathbf{l}_{k}\right|^{2}\left(1-\cos^{2}\phi_{ik}\right),
\]
or: 
\begin{equation}
\left|\mathbf{l}_{i}\wedge\mathbf{l}_{k}\right|=\left|\mathbf{l}_{i}\right|\left|\mathbf{l}_{k}\right|\sin\phi_{ik}.\label{eq:7}
\end{equation}

\subparagraph{Step 3:}

Inserting (\ref{eq:6}) and using (\ref{eq:7}) to (\ref{eq:2}),
we obtain:
\begin{eqnarray*}
g\left(\mathbf{l}_{i}\wedge\mathbf{l}_{k},\mathbf{l}_{j}\wedge\mathbf{l}_{k}\right) & = & \left|\mathbf{l}_{i}\wedge\mathbf{l}_{k}\right|\left|\mathbf{l}_{j}\wedge\mathbf{l}_{k}\right|\cos\theta_{ij,k}\\
\left|\mathbf{l}_{i}\right|\left|\mathbf{l}_{k}\right|^{2}\left|\mathbf{l}_{j}\right|\left(\cos\phi_{ij}-\cos\phi_{ik}\cos\phi_{kj}\right) & = & \left|\mathbf{l}_{i}\right|\left|\mathbf{l}_{k}\right|^{2}\left|\mathbf{l}_{j}\right|\sin\phi_{ik}\sin\phi_{kj}\cos\theta_{ij,k}
\end{eqnarray*}
or:
\begin{equation}
\cos\theta_{ij,k}=\frac{\cos\phi_{ij}-\cos\phi_{ik}\cos\phi_{kj}}{\sin\phi_{ik}\sin\phi_{kj}}.\label{eq:8-1}
\end{equation}

\section{Coordinate-free variables for tetrahedron and 4-simplex}

\subsection{Tetrahedron}

To proof that a set of variables describe uniquely the geometry of
a tetrahedron is to show that a set of six unique length of edges
can be obtained by a corresponding transformation. Let us start with
$\left\{ \left|\mathbf{a}_{i}\right|,\left|\mathbf{a}_{j}\right|,\left|\mathbf{a}_{k}\right|,\theta_{ij},\theta_{ik},\theta_{jk}\right\} .$
See FIG. 14. 
\begin{figure}[h]
\begin{centering}
\includegraphics[scale=0.85]{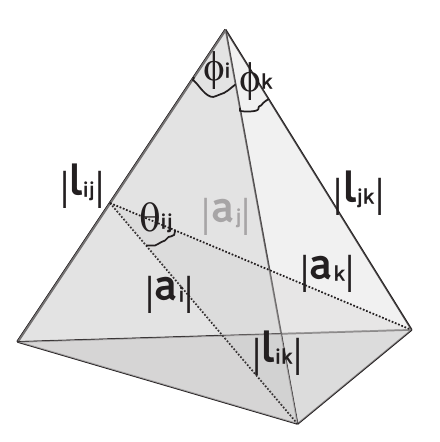}\caption{Terminologies of the variables of a tetrahedron.}
\par\end{centering}
\end{figure}
Using the inverse dihedral angle relation as follows:
\begin{equation}
\cos\phi_{i}=\frac{\cos\theta_{jk}-\cos\theta_{ij}\cos\theta_{ik}}{\sin\theta_{ij}\sin\theta_{ik}},\label{eq:b0}
\end{equation}
we could obtain the single-value 2D dihedral angles $\left\{ \phi_{i},\phi_{j},\phi_{k}\right\} $
from $\left\{ \theta_{ij},\theta_{ik},\theta_{jk}\right\} $ in the
range of $0<\phi<\pi$. The next step is to solve the following system
of linear equations:
\begin{equation}
\left|\mathbf{a}_{i}\right|=\frac{1}{2}\left|\mathbf{l}_{ij}\right|\underset{\left|\mathbf{h}_{2}\right|}{\underbrace{\left|\mathbf{l}_{ik}\right|\sin\phi_{i},}}\label{eq:b1}
\end{equation}
for $\left\{ \left|\mathbf{l}_{ij}\right|,\left|\mathbf{l}_{ik}\right|,\left|\mathbf{l}_{jk}\right|\right\} .$
(\ref{eq:b1}) is only the area formula of a triangle. Having three
lengths of edges of the tetrahedron (see FIG. 14), we could obtain
the remaining three edges from the law of cosine:
\begin{equation}
\left|\mathbf{L}_{i}\right|^{2}=\left|\mathbf{l}_{ij}\right|^{2}+\left|\mathbf{l}_{ik}\right|^{2}-2\left|\mathbf{l}_{ij}\right|\left|\mathbf{l}_{ik}\right|\cos\phi_{i}.\label{eq:b4}
\end{equation}
Since we assume the length is positive definite, then we obtain single
value of $\left\{ \left|\mathbf{L}_{i}\right|,\left|\mathbf{L}_{j}\right|,\left|\mathbf{L}_{k}\right|\right\} .$
Therefore, the map between $\left\{ \left|\mathbf{a}_{i}\right|,\left|\mathbf{a}_{j}\right|,\left|\mathbf{a}_{k}\right|,\theta_{ij},\theta_{ik},\theta_{jk}\right\} $
and $\left\{ \left|\mathbf{l}_{ij}\right|,\left|\mathbf{l}_{ik}\right|,\left|\mathbf{l}_{jk}\right|,\left|\mathbf{L}_{i}\right|,\left|\mathbf{L}_{j}\right|,\left|\mathbf{L}_{k}\right|\right\} $
is one-to-one.

Let us proof the uniqueness of the next choice of variables: $\left\{ \left|\mathbf{a}_{i}\right|,\left|\mathbf{a}_{j}\right|,\left|\mathbf{a}_{k}\right|,\left|\mathbf{a}_{ijk}\right|,\theta_{ij},\theta_{ik}\right\} .$
The easiest way is to obtain the remaining 3D dihedral angle $\theta_{jk}.$
To do this we use the closure condition of the tetrahedron (\ref{eq:3.12}) to obtain:
\begin{equation}
\left|\mathbf{a}_{ijk}\right|^{2}=\left|\mathbf{a}_{i}\right|^{2}+\left|\mathbf{a}_{j}\right|^{2}+\left|\mathbf{a}_{k}\right|^{2}-2\left(\left|\mathbf{a}_{i}\right|\left|\mathbf{a}_{j}\right|\cos\theta_{ij}+\left|\mathbf{a}_{i}\right|\left|\mathbf{a}_{k}\right|\cos\theta_{ik}+\left|\mathbf{a}_{j}\right|\left|\mathbf{a}_{k}\right|\cos\theta_{jk}\right).\label{eq:b2}
\end{equation}
Solving (\ref{eq:b2}) for $\theta_{jk}$ (which is unique since $0<\theta_{jk}<\pi$)
gives the information of $\left\{ \left|\mathbf{a}_{i}\right|,\left|\mathbf{a}_{j}\right|,\left|\mathbf{a}_{k}\right|,\theta_{ij},\theta_{ik},\theta_{jk}\right\} $,
and the next step to obtain the edges lengths is already obvious from
the previous proof. Therefore, the map between $\left\{ \left|\mathbf{a}_{i}\right|,\left|\mathbf{a}_{j}\right|,\left|\mathbf{a}_{k}\right|,\left|\mathbf{a}_{ijk}\right|,\theta_{ij},\theta_{ik}\right\} $
and $\left\{ \left|\mathbf{l}_{ij}\right|,\left|\mathbf{l}_{ik}\right|,\left|\mathbf{l}_{jk}\right|,\left|\mathbf{L}_{i}\right|,\left|\mathbf{L}_{j}\right|,\left|\mathbf{L}_{k}\right|\right\} $
is one-to-one.

The last proof is for the choice of variables $\left\{ \left|\mathbf{v}\right|,\left|\mathbf{a}_{i}\right|,\left|\mathbf{a}_{j}\right|,\theta_{ij},\theta_{ik},\theta_{jk}\right\} .$
For this case, the easiest way is to obtain the single-value 2D dihedral
angles $\left\{ \phi_{i},\phi_{j},\phi_{k}\right\} $ from $\left\{ \theta_{ij},\theta_{ik},\theta_{jk}\right\} $
using (\ref{eq:b0}). The volume of a tetrahedron can be obtained
from the relation as follows:
\begin{equation}
\left|\mathbf{v}\right|=\frac{2}{3}\left|\mathbf{a}_{i}\right|\left|\mathbf{a}_{j}\right|\frac{\sin\theta_{ij}}{\left|\mathbf{l}_{ij}\right|},\label{eq:b3}
\end{equation}
where we can solve (\ref{eq:b3}) for a unique $\left|\mathbf{l}_{ij}\right|.$
With $\left|\mathbf{l}_{ij}\right|$ known, the other edges length
$\left|\mathbf{l}_{ik}\right|$ and $\left|\mathbf{l}_{jk}\right|$
can be obtained from (\ref{eq:b1}). Having the information of $\left\{ \phi_{i},\phi_{j},\phi_{k},\left|\mathbf{l}_{ij}\right|,\left|\mathbf{l}_{ik}\right|,\left|\mathbf{l}_{jk}\right|\right\} $
(see FIG. B1), we could obtain the remaining three edges from the
law of cosine (\ref{eq:b4}). Therefore, the map between $\left\{ \left|\mathbf{v}\right|,\left|\mathbf{a}_{i}\right|,\left|\mathbf{a}_{j}\right|,\theta_{ij},\theta_{ik},\theta_{jk}\right\} $
and $\left\{ \left|\mathbf{l}_{ij}\right|,\left|\mathbf{l}_{ik}\right|,\left|\mathbf{l}_{jk}\right|,\left|\mathbf{L}_{i}\right|,\left|\mathbf{L}_{j}\right|,\left|\mathbf{L}_{k}\right|\right\} $
is one-to-one.

\subsection{4-simplex}

The geometry of a 4-simplex is described uniquely by their ten edges.
As a starting point, let us study the variables of a 4-simplex as
follows. Let the four edges of 4-simplex be denoted by $\left|\mathbf{l}_{a}\right|$,
.., $\left|\mathbf{l}_{d}\right|$, and the six remaining edges be
denoted by $\left|\mathbf{l}_{ab}\right|,$ .., $\left|\mathbf{l}_{cd}\right|$.
The 2D angle between two edges, say, $\left|\mathbf{l}_{a}\right|$
and $\left|\mathbf{l}_{b}\right|$, is denoted by $\phi_{ab}$. See
FIG. 15. 
\begin{figure}[h]
\centering{}\includegraphics[scale=0.85]{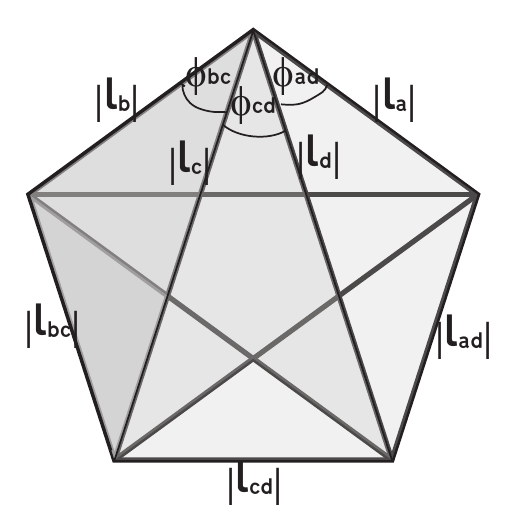}\caption{Terminologies of the edges of a tetrahedron.}
\end{figure}
 Now let us denote the four tetrahedra of the 4-simplex by $\left|\mathbf{v}_{i}\right|$,
$i=1,..,4$, and the 4D angles between two adjacent tetrahedra $\left|\mathbf{v}_{i}\right|$
and $\left|\mathbf{v}_{j}\right|$ as $\varphi_{ij}$, see FIG. 16.
\begin{figure}[h]
\centering{}\includegraphics[scale=0.85]{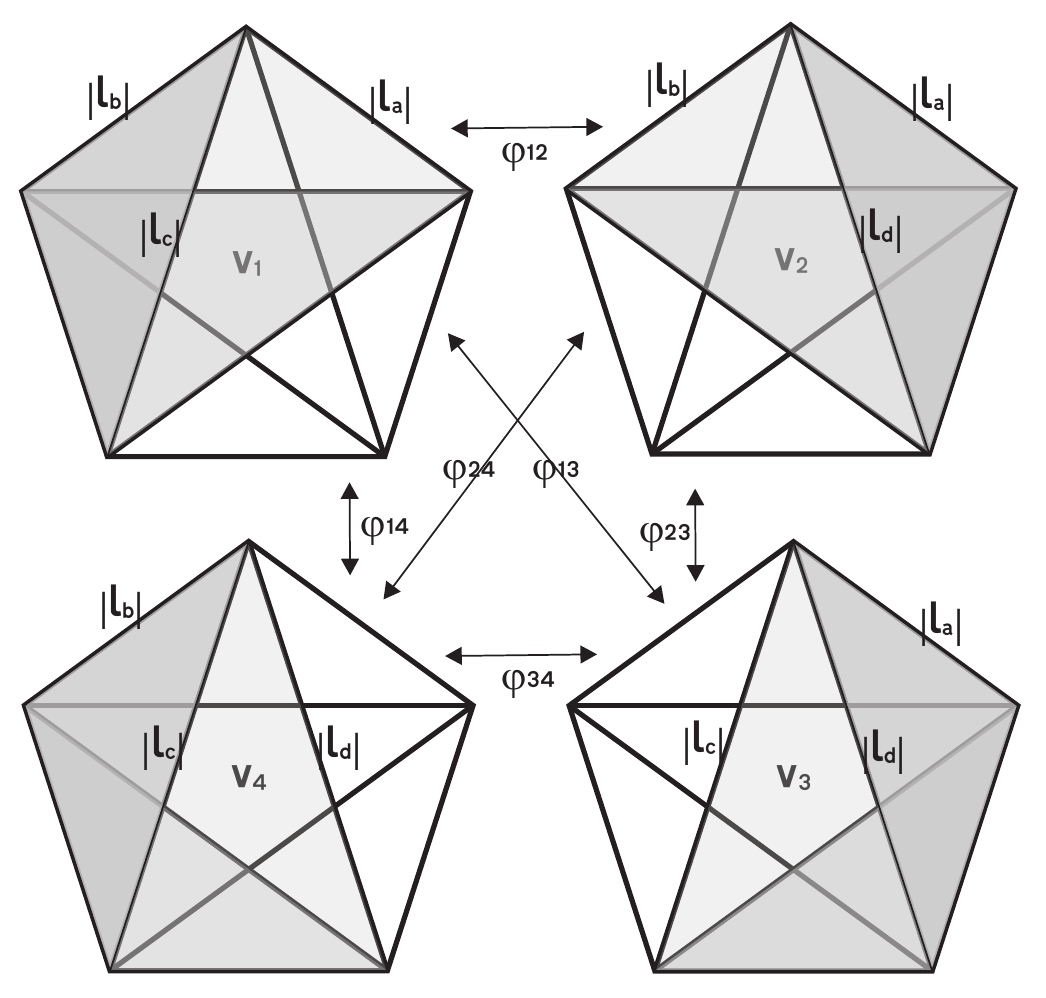}\caption{Four tetrahedra (pictured in gray) of the 4-simplex. Each tetrahedron
consists six edges; for example, tetrahedron $\left|\mathbf{v}_{1}\right|$
is built from edges $\left|\mathbf{l}_{a}\right|$, $\left|\mathbf{l}_{b}\right|$,
$\left|\mathbf{l}_{c}\right|$, $\left|\mathbf{l}_{ab}\right|$, $\left|\mathbf{l}_{ac}\right|$,
and $\left|\mathbf{l}_{bc}\right|$. The 4D angle between two tetrahedra
is denoted by $\varphi_{ij}$; for example, the 4D angle between tetrahedron
$\left|\mathbf{v}_{1}\right|$ and $\left|\mathbf{v}_{2}\right|$
is $\varphi_{12}$.}
\end{figure}
 Let the 'base' tetrahedron denoted by $\left|\mathbf{v}_{1234}\right|$,
as illustrated in FIG. 17. 
\begin{figure}[h]
\centering{}\includegraphics[scale=0.85]{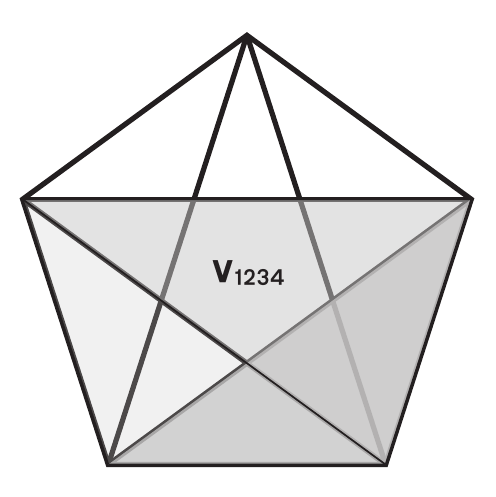}\caption{The 'base' tetrahedron.}
\end{figure}

The 4D angles $\varphi_{ij}$ are located on a hinge, which is a triangle.
See FIG. 18 for an example. 
\begin{figure}[h]

\begin{centering}
\includegraphics[scale=0.85]{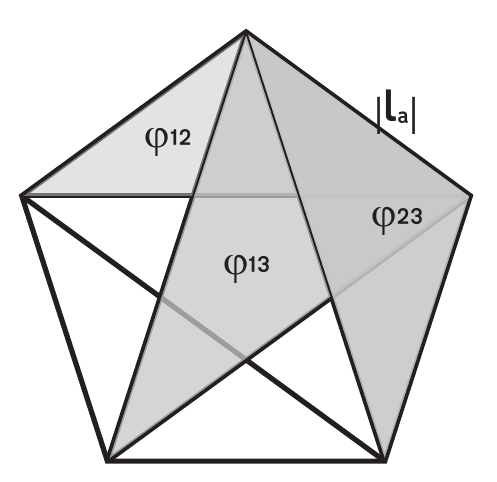}\caption{The 4D angles $\varphi_{12}$, $\varphi_{13}$, and $\varphi_{23}$
are located on triangles. These triangles meet on a common edge $\left|\mathbf{l}_{a}\right|.$ }

\par\end{centering}

\end{figure}
The 3D angle $\theta_{i,n}$ is defined as the angle between two triangles
which contains $\varphi_{ij}$ and $\varphi_{ik}$, meeting on a common
edge $\left|\mathbf{l}_{n}\right|$, see FIG. 19 for examples. 
\begin{figure}[h]

\begin{centering}
\includegraphics[scale=0.85]{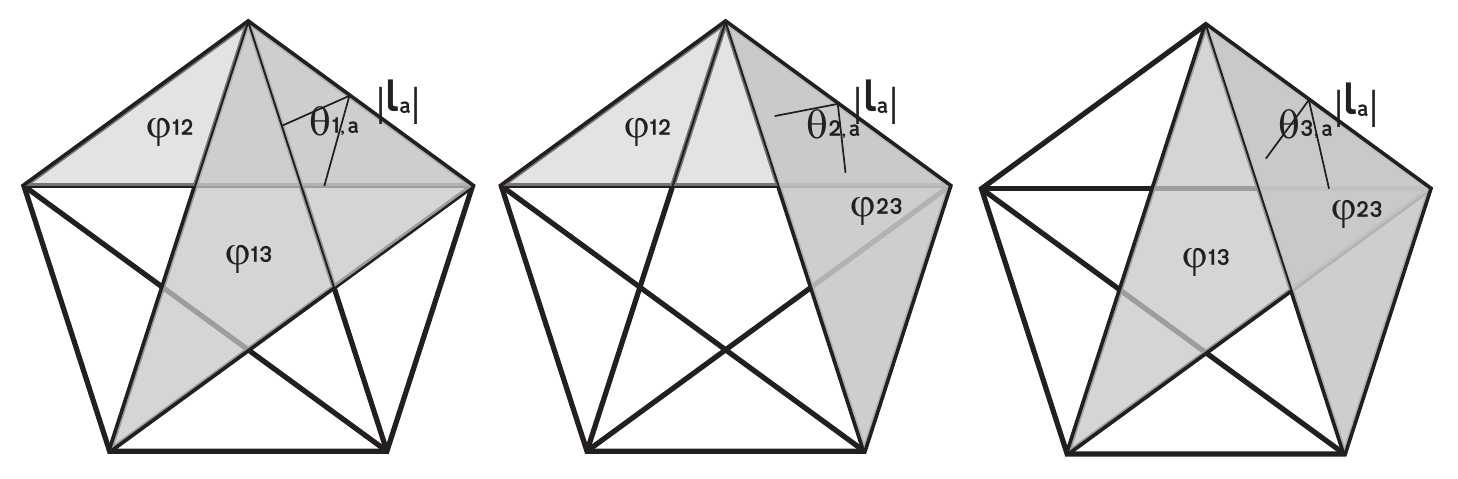}\caption{The three triangles where the 4D angles $\left\{ \varphi_{12},\varphi_{13},\varphi_{23}\right\} $
are located, defines the 3D angles $\left\{ \theta_{1,a},\theta_{2,a},\theta_{3,a}\right\} $
located on $\left|\mathbf{l}_{a}\right|.$ }

\par\end{centering}

\end{figure}

Now, we can start to proof that a set of four 3-volumes of tetrahedra
with six 4D dihedral angles $\left\{ \left|\mathbf{v}_{i}\right|,\varphi_{ij}\right\} $,
$i,j=1,..,4$, $i<j$, describe uniquely the geometry of a 4-simplex.
Using the (3+1) dihedral angle relation:
\[
\cos\theta_{i,n}=\frac{\cos\varphi_{jk}-\cos\varphi_{ij}\cos\varphi_{ik}}{\sin\varphi_{ij}\sin\varphi_{ik}},
\]
we could obtain the 3D angles as follows:
\begin{eqnarray*}
\left\{ \theta_{1,a},\theta_{2,a},\theta_{3,a}\right\} \, & \textrm{from} & \,\left\{ \varphi_{12},\varphi_{13},\varphi_{23}\right\} ,\\
\left\{ \theta_{1,b},\theta_{2,b},\theta_{4,b}\right\} \, & \textrm{from} & \,\left\{ \varphi_{12},\varphi_{14},\varphi_{24}\right\} ,\\
\left\{ \theta_{1,c},\theta_{3,c},\theta_{4,c}\right\} \, & \textrm{from} & \,\left\{ \varphi_{13},\varphi_{14},\varphi_{34}\right\} ,\\
\left\{ \theta_{2,d},\theta_{3,d},\theta_{4,d}\right\} \, & \textrm{from} & \,\left\{ \varphi_{23},\varphi_{24},\varphi_{34}\right\} .
\end{eqnarray*}
These 3D angles are in the range of $0<\theta_{i,n}<\pi$ and therefore,
unique. Moreover, we classify the 3D dihedral angles according to
their tetrahedra: $\theta_{i,n}$ for any $n$ belongs to tetrahedron
$\left|\mathbf{v}_{i}\right|,$ see FIG. 20 for example. 
\begin{figure}[h]
\begin{centering}
\includegraphics[scale=0.85]{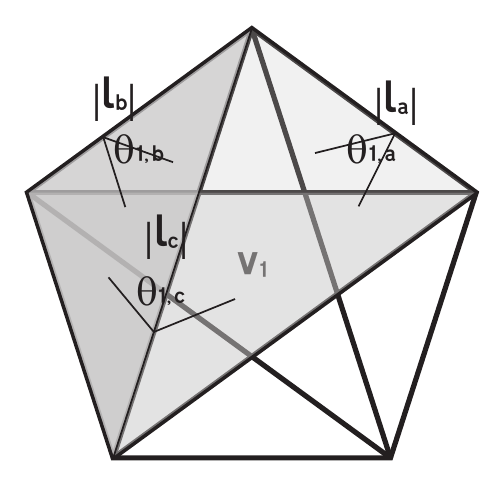}\caption{The 3D angles $\theta_{1,a}$, $\theta_{1,b},$ and $\theta_{1,c},$
belongs to tetrahedron $\left|\mathbf{v}_{1}\right|,$ so that they
satisfy the (2+1) dihedral angles relation with the 2D angles $\phi$. }

\par\end{centering}

\end{figure}
 Adapting the (2+1) dihedral angle relation (\ref{eq:b1}) with the
previous terminologies of the 4-simplex as follows:
\[
\cos\phi_{ij}=\frac{\cos\theta_{l,k}-\cos\theta_{l,i}\cos\theta_{l,j}}{\sin\theta_{l,i}\sin\theta_{l,j}},
\]
we could obtain six 2D dihedral angles:
\begin{eqnarray*}
\left\{ \phi_{ab},\phi_{ac},\phi_{bc}\right\} \, & \textrm{from} & \,\left\{ \theta_{1,a},\theta_{1,b},\theta_{1,c}\right\} ,\\
\left\{ \phi_{ab},\phi_{ad},\phi_{bd}\right\} \, & \textrm{from} & \,\left\{ \theta_{2,a},\theta_{2,b},\theta_{2,d}\right\} ,\\
\left\{ \phi_{ac},\phi_{ad},\phi_{cd}\right\} \, & \textrm{from} & \,\left\{ \theta_{3,a},\theta_{3,c},\theta_{3,d}\right\} ,\\
\left\{ \phi_{bc},\phi_{bd},\phi_{cd}\right\} \, & \textrm{from} & \,\left\{ \theta_{4,b},\theta_{4,c},\theta_{4,d}\right\} .
\end{eqnarray*}
Having the information of six dihedral angles $\left\{ \phi_{ab},\phi_{ac},\phi_{ad},\phi_{bc},\phi_{bd},\phi_{cd}\right\} $
and the four volumes of tetrahedra $\left\{ \left|\mathbf{v}_{1}\right|,\left|\mathbf{v}_{2}\right|,\left|\mathbf{v}_{3}\right|,\left|\mathbf{v}_{4}\right|\right\} $,
we could solve the tetrahedron volume formula:
\begin{eqnarray}
\left|\mathbf{v}_{1}\right| & = & \frac{1}{3}\left|\mathbf{l}_{a}\right|\left|\mathbf{l}_{b}\right|\left|\mathbf{l}_{c}\right|\sqrt{1+2\cos\phi_{ab}\cos\phi_{ac}\cos\phi_{bc}-\left(\cos^{2}\phi_{ab}+\cos^{2}\phi_{ac}+\cos^{2}\phi_{bc}\right)},\label{eq:b7}\\
\left|\mathbf{v}_{2}\right| & = & \frac{1}{3}\left|\mathbf{l}_{a}\right|\left|\mathbf{l}_{b}\right|\left|\mathbf{l}_{d}\right|\sqrt{1+2\cos\phi_{ab}\cos\phi_{ad}\cos\phi_{bd}-\left(\cos^{2}\phi_{ab}+\cos^{2}\phi_{ad}+\cos^{2}\phi_{bd}\right)},\label{eq:b8}\\
\left|\mathbf{v}_{3}\right| & = & \frac{1}{3}\left|\mathbf{l}_{a}\right|\left|\mathbf{l}_{c}\right|\left|\mathbf{l}_{d}\right|\sqrt{1+2\cos\phi_{ac}\cos\phi_{ad}\cos\phi_{cd}-\left(\cos^{2}\phi_{ac}+\cos^{2}\phi_{ad}+\cos^{2}\phi_{cd}\right)},\label{eq:b9}\\
\left|\mathbf{v}_{4}\right| & = & \frac{1}{3}\left|\mathbf{l}_{b}\right|\left|\mathbf{l}_{c}\right|\left|\mathbf{l}_{d}\right|\sqrt{1+2\cos\phi_{bc}\cos\phi_{bd}\cos\phi_{cd}-\left(\cos^{2}\phi_{bc}+\cos^{2}\phi_{bd}+\cos^{2}\phi_{cd}\right)},\label{eq:b10}
\end{eqnarray}
for the four edges $\left\{ \left|\mathbf{l}_{a}\right|,\left|\mathbf{l}_{b}\right|,\left|\mathbf{l}_{c}\right|,\left|\mathbf{l}_{d}\right|\right\} $,
which is unique for each length.

For the last step, having the information of $\left\{ \left|\mathbf{l}_{a}\right|,\left|\mathbf{l}_{b}\right|,\left|\mathbf{l}_{c}\right|,\left|\mathbf{l}_{d}\right|\right\} $
and $\left\{ \phi_{ab},..,\phi_{cd}\right\} $, we could obtain six
remaining edges of the 4-simplex $\left\{ \left|\mathbf{l}_{ab}\right|,..,\left|\mathbf{l}_{cd}\right|\right\} $
from the law of cosine (\ref{eq:b4}). Therefore, the map between
$\left\{ \left|\mathbf{v}_{i}\right|,\varphi_{ij}\right\} $, $i,j=1,..,4$,
$i<j$, and ten edges $\left\{ \left|\mathbf{l}_{a}\right|,..,\left|\mathbf{l}_{d}\right|,\left|\mathbf{l}_{ab}\right|,..,\left|\mathbf{l}_{ca}\right|\right\} $
is one-to-one.

Let us proof the uniqueness of the next choice of variables of a 4-simplex:
the 4-volume of the 4-simplex, three 3-volumes of tetrahedra, and
six 4D dihedral angles $\left\{ \left|\mathbf{s}\right|,\left|\mathbf{v}_{i}\right|,\varphi_{jk}\right\} $,
$i=1,2,3$, $j,k=1,..,4,$ $j<k$. The 'volume' (called as area) of
a 2D triangle, can be obtained from a conventional area formula (\ref{eq:b1}),
which is half of the 'base' edge times its 'height' $\left|\mathbf{h}_{2}\right|$.
An analog formula is also valid for tetrahedron and 4-simplex \cite{key-3.29a}. See
FIG. 21. 
\begin{figure}[h]
\begin{centering}
\includegraphics[scale=0.85]{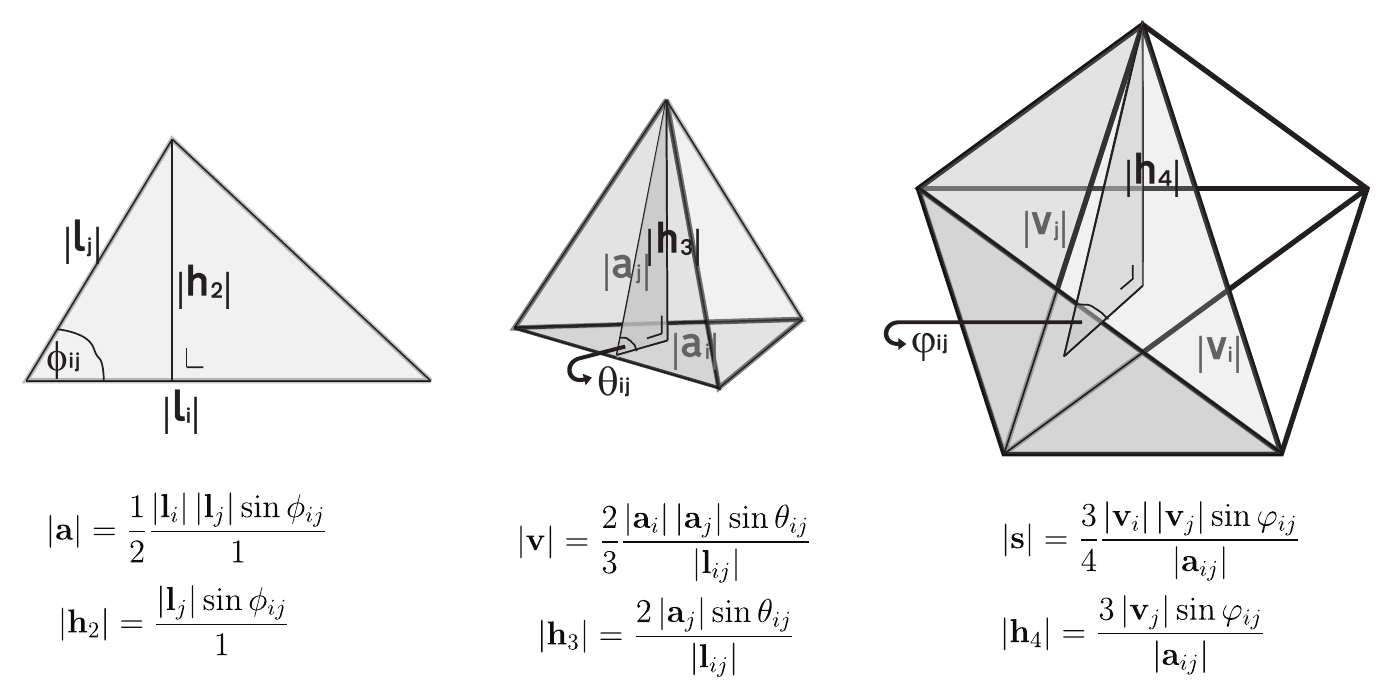}\caption{For a triangle, we have a conventional area formula, which is half
of the base $\left|\mathbf{l}_{i}\right|$ times its height $\left|\mathbf{h}_{2}\right|.$
For a tetrahedron, a similar formula is valid for the volume, which
is one-third of the 'base', which is now the area of a triangle $\left|\mathbf{a}_{i}\right|$,
times its height $\left|\mathbf{h}_{3}\right|.$ Moreover, we could
obtain an analog for a 4-simplex: a quarter of the 'base', which,
recursively, is the volume of a tetrahedron, times its height $\left|\mathbf{h}_{4}\right|.$ }

\par\end{centering}

\end{figure}

The volume of a 4-simplex could be written as:
\begin{equation}
\left|\mathbf{s}\right|=\frac{3}{4}\frac{\left|\mathbf{v}_{i}\right|\left|\mathbf{v}_{j}\right|\sin\varphi_{ij}}{\left|\mathbf{a}_{ij}\right|},\label{eq:b5}
\end{equation}
given two volumes of tetrahedra $\left|\mathbf{v}_{i}\right|$ and
$\left|\mathbf{v}_{j}\right|$ meeting on a common triangle $\left|\mathbf{a}_{ij}\right|.$
This is a 4-dimensional analog to (\ref{eq:b1}) and (\ref{eq:b3}).
Let us choose $i=1$ and $j=2,$ so that 
\[
\left|\mathbf{s}\right|=\frac{3}{4}\frac{\left|\mathbf{v}_{1}\right|\left|\mathbf{v}_{2}\right|\sin\varphi_{12}}{\left|\mathbf{a}_{12}\right|},
\]
see FIG. 16. It is clear that we can write $\left|\mathbf{a}_{12}\right|$
as:
\[
\left|\mathbf{a}_{12}\right|=\frac{1}{2}\left|\mathbf{l}_{a}\right|\left|\mathbf{l}_{b}\right|\sin\phi_{ab},
\]
so that we have:
\begin{equation}
\left|\mathbf{s}\right|=\frac{3}{2}\frac{\left|\mathbf{v}_{1}\right|\left|\mathbf{v}_{2}\right|}{\left|\mathbf{l}_{a}\right|\left|\mathbf{l}_{b}\right|}\frac{\sin\varphi_{12}}{\sin\phi_{ab}}.\label{eq:b6}
\end{equation}

Our choice of variables of the 4-simplex are $\left\{ \left|\mathbf{s}\right|,\left|\mathbf{v}_{i}\right|,\varphi_{jk}\right\} $,
$i=1,2,3$, $j,k=1,..,4,$ $j<k$. From the six 4D angles $\varphi_{jk}$,
$j,k=1,..,4,$ we could obtain six unique 2D angles $\left\{ \phi_{ab},..,\phi_{cd}\right\} $
by the previous derivation. But this time, we only have the information
of three volumes of tetrahedra, instead of four, so we only have equation
(\ref{eq:b7}), (\ref{eq:b8}) and (\ref{eq:b9}). To solve these
equations for four edges $\left\{ \left|\mathbf{l}_{a}\right|,\left|\mathbf{l}_{b}\right|,\left|\mathbf{l}_{c}\right|,\left|\mathbf{l}_{d}\right|\right\} $,
we need one more equation, which comes from the 4-simplex volume relation
(\ref{eq:b6}). Having unique value of $\left\{ \left|\mathbf{l}_{a}\right|,\left|\mathbf{l}_{b}\right|,\left|\mathbf{l}_{c}\right|,\left|\mathbf{l}_{d}\right|\right\} $,
with their 2D angles, we could obtain six remaining edges of the 4-simplex
$\left\{ \left|\mathbf{l}_{ab}\right|,..,\left|\mathbf{l}_{cd}\right|\right\} $
from the law of cosine (\ref{eq:b4}) just as the previous derivation.
Therefore, the map between $\left\{ \left|\mathbf{s}\right|,\left|\mathbf{v}_{i}\right|,\varphi_{jk}\right\} $,
$i=1,2,3$, $j,k=1,..,4,$ $j<k$, and ten edges $\left\{ \left|\mathbf{l}_{a}\right|,..,\left|\mathbf{l}_{d}\right|,\left|\mathbf{l}_{ab}\right|,..,\left|\mathbf{l}_{ca}\right|\right\} $
is one-to-one.

\bibliographystyle{apsrev4-1}

   \bibliography{library}\end{document}